\documentclass[sigconf,10pt]{acmart}    
\usepackage{multirow}
\usepackage{tikz}
\usepackage{makecell} 
\usepackage{lscape} 
\usepackage{siunitx}
\usepackage{bm}     
\usepackage{makecell}
\usepackage{pifont}
\usepackage{url}
\usepackage{siunitx} 
\usepackage{xspace}
\usepackage{algorithm}
\usepackage{algorithmic}
\usepackage[caption=false,font=normalsize,labelfont=sf,textfont=sf]{subfig}  
 
\definecolor{codecolor}{rgb}{0.902,0.518,0.259}
\newcommand{\systemname}{{UrgenGo}\xspace}
  
\newcommand{\buffername}{{AKB}\xspace}

\AtBeginDocument{%
  }

\setcopyright{acmlicensed}
\copyrightyear{2025}
\acmYear{2025}
\acmDOI{xx.xxxx/xxxxxxx.xxxxxxx} 
\acmISBN{xxx-x-xxxx-xxxx-x/xx/xx}

\begin{document}
 
\title{\systemname: Urgency-Aware Transparent GPU Kernel Launching for Autonomous Driving}   
 
\renewcommand{\shorttitle}{ \systemname: Urgency-Aware Kernel Launching}

\begin{sloppypar}
\begin{abstract}
The rapid advancements in autonomous driving have introduced increasingly complex, real-time GPU-bound tasks critical for reliable vehicle operation. 
However, the proprietary nature of these autonomous systems and closed-source GPU drivers hinder fine-grained control over GPU executions, often resulting in missed deadlines that compromise vehicle performance.
To address this, we present \systemname, a non-intrusive, urgency-aware GPU scheduling system that operates without access to application source code. \systemname implicitly prioritizes GPU executions through transparent kernel launch manipulation, employing task-level stream binding, delayed kernel launching, and batched kernel launch synchronization. 
We conducted extensive real-world evaluations in collaboration with a self-driving startup, developing 11 GPU-bound task chains for a realistic autonomous navigation application and implementing our system on a self-driving bus. 
Our results show a significant 61\% reduction in the overall deadline miss ratio, compared to the state-of-the-art GPU scheduler that requires source code modifications. 
\end{abstract}



\author{Hanqi Zhu$^\dagger$, Wuyang Zhang$^\dagger$, Xinran Zhang$^\dagger$, Ziyang Tao$^\dagger$, Xinrui Lin$^\dagger$, Yu Zhang$^{\dagger, \diamondsuit}$, Jianmin Ji$^{\dagger, \diamondsuit}$, Yanyong Zhang$^{\dagger, \diamondsuit}$\\
\small $^\dagger$ University of Science and Technology of China \quad 
\small $^\diamondsuit$ Institute of Artificial Intelligence, Hefei Comprehensive National Science Center \\
\vspace{-5pt}
\{zhuhanqi, zxrr, ziyang, xinruilin\}@mail.ustc.edu.cn, \{wuyangz, yuzhang, jianmin, yanyongz\}@ustc.edu.cn
}

\renewcommand{\shortauthors}{Zhu et al.} 

\maketitle
 
\section{Introduction  }   
    Autonomous driving has emerged as a transformative technology with the potential to revolutionize transportation by enhancing safety, improving efficiency, and simplifying daily life. 
    However, before these promises can be fully realized, key technical challenges need to be addressed--chief among them is meeting the processing demands and real-time deadlines of various tasks within autonomous driving systems. 
    As vehicle sensors increase in resolution, deep learning models grow in complexity, and self-driving cars expand in services, the computational requirements outpace the capabilities of in-vehicle computing platforms.
    This growing disparity makes it increasingly difficult to meet the time constraints of the applications, thereby undermining system functionality~\cite{ApolloTiming,liu2020computing}. 
 
\begin{figure}[t]
        \centering 
        \includegraphics[width=.97\linewidth]{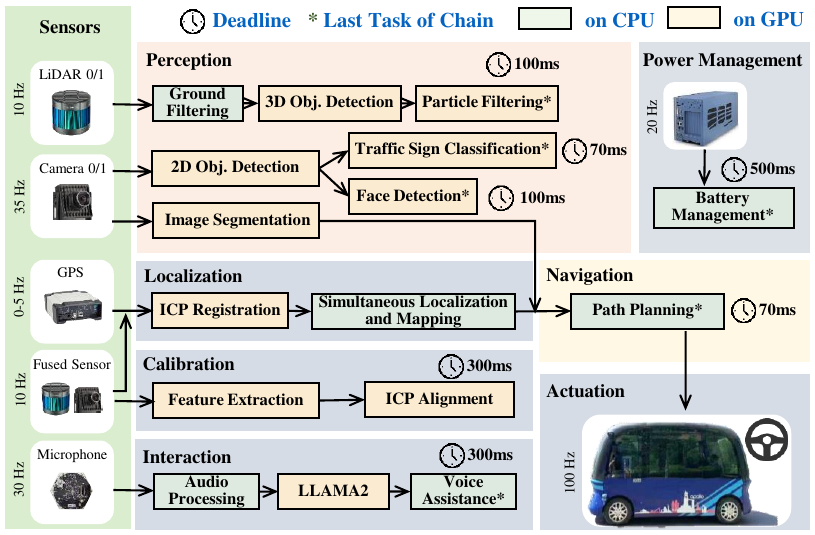} 
        \vspace{-12pt}
        \caption{ \small An example with 4 different sensors and 9 task chains. The last task of a chain is marked with a star. 
        } \label{fig:workflow1} 
        \vspace{-16pt} 
    \end{figure}

    Fig.~\ref{fig:workflow1} shows a representative yet simplified workflow of an autonomous driving system. 
    Despite the dynamic environment, the system follows a relatively fixed workflow at the task level, typically involving tasks related to  perception~\cite{vpfnet,shi2022vips}, navigation~\cite{he2023vi}, localization~\cite{map++,sie2024radarize}, calibration~\cite{calibration}, etc.
    These tasks are organized into task chains, with each chain initiated by periodically arriving sensor data and tasks triggered sequentially in a data-driven manner. 
    Each chain usually adheres to a predefined end-to-end deadline. 
    For example, among the 9 task chains shown in Fig.~\ref{fig:workflow1}, some (e.g., path planning) operate under tight deadlines, while others (e.g., online calibration) can tolerate longer delays.   
    
    In this workflow, tasks that leverage GPU processing often become the most time-consuming components (e.g., those highlighted in yellow in Fig.~\ref{fig:workflow1}), with a single task involving hundreds of GPU kernels.   
    Effectively scheduling these kernels across the CPU and GPU is critical, especially in systems with multiple chains and varying timing requirements. 
    This challenge has received attention in robotics~\cite{enright2024paam,ling2022blastnet,zhao2023miriam}, AR/VR services~\cite{yi2020heimdall}, and other cyber-physical systems~\cite{han2024pantheon,s3dnn}. 
    Existing techniques generally fall into three categories: (1) GPU time-sharing, (2) GPU space-sharing, and (3) kernel launch prioritization.  
    For instance,  
    {Heimdall~\cite{yi2020heimdall} and dCUDA~\cite{ccuda} enable time-sharing by partitioning applications into groups and executing them in distinct time slices.  cCUDA~\cite{ccuda} enhances GPU concurrency through space sharing by splitting kernels into sub-kernels.  
    PAAM~\cite{enright2024paam} and DART~\cite{xiang2019pipelined} implement kernel launch prioritization by binding tasks to CUDA streams based on predefined criticality.} 
 
    However, most of these methods require modifying the application \emph{source code}, which is impractical for autonomous driving systems relying on closed-source, third-party libraries. 
    For example, many original equipment manufacturers (OEMs) utilize vendor-optimized frameworks like TensorRT~\cite{nvidia_tensorrt} for kernel acceleration, along with algorithm libraries~\cite{lidar_ai,isaac_slam} provided by tier 1/2 suppliers, to minimize costs.
    While overall task topologies are known, task implementations remain hidden within proprietary frameworks. 
  
    Furthermore, many existing GPU prioritization works rely heavily on \emph{static} priorities. 
    For example, PAAM~\cite{enright2024paam} and DART~\cite{xiang2019pipelined} assign priorities based on the criticality of each task chain, as predefined by the application.
    While static prioritization provides better determinism, it becomes less practical in autonomous driving, where multiple chains may share similar levels of criticality. In such cases, dynamic prioritization, which accounts for real-time factors such as urgency and deadline, is more appropriate. Specifically, we determine a task's dynamic priority based on its urgency~\cite{stewart1991real}, defined as inversely proportional to its laxity. 
    However, this introduces new challenges, including (1) how to accurately estimate the execution status and urgency of GPU kernels from the CPU with minimal overhead, and (2) how to utilize the dynamic urgency to guide GPU kernel launches such that their execution aligns with corresponding urgency levels.   
 
    In response to these challenges, we present \systemname, an \textit{urgency-centric, transparent} task scheduling system, designed for \textit{closed-source multi-task-chain} applications with mixed real-time constraints and hybrid CPU-GPU execution.  
    The key idea behind \systemname is to dynamically schedule GPU kernels based on their changing urgency level -- e.g., tasks become more urgent as deadlines approach with insufficient progress, or less urgent after missing deadlines. This is achieved   
    through non-intrusive CUDA kernel launch manipulation: \systemname intercepts CUDA function calls by replacing the default CUDA dynamic library with a custom variant, allowing GPU execution \textit{prioritization} without modifying the application source code at all. \systemname provides refined control over kernel launch \textit{priority, timing, and synchronization} within a unified framework.  
    In particular, it introduces task-level CUDA stream binding to minimize priority collisions, delayed kernel launching to prevent GPU execution collisions, and batched kernel launching to balance task urgency tracking with execution efficiency.
    Overall, \systemname is a plugin-compatible framework that enables urgency-aware scheduling for autonomous driving, addressing priority-aware GPU kernel launch challenges and optimizing scheduling for real-time closed-source applications.
 
    To proceed with a realistic evaluation, we collaborated with a pioneering self-driving startup and selected 11 representative AI-driven task chains from an autonomous navigation application. {These task chains span 6 service domains: 3D perception, 2D perception, localization, navigation, calibration, and interaction.} We implemented the system on a self-driving bus and collected data over two weeks in a 480,000-square-meter campus with diverse conditions.
    This paper makes four primary contributions: 
       \begin{enumerate}
        \vspace{-2pt}\item We propose a transparent, urgency-aware GPU scheduling framework that operates without source code access. To the best of our knowledge, we are the first to manipulate all key kernel launch parameters -- priority, timing, and synchronization mode -- specifically for autonomous driving systems that rely on closed-source libraries with time constraints. 
        \vspace
        {-0pt}\item We design and implement key system components: a transparent kernel launch manipulation module that integrates priority, timing, and synchronization in a unified framework, and an urgency-centric scheduler that manages chain-level CPU prioritization, task-level CUDA stream binding, kernel-level delayed launching, and kernel batch-level synchronization. Together, these modules enable fine-grained real-time GPU scheduling under dynamic workloads.

        \vspace{-0pt}\item We conducted a comprehensive real-world evaluation using 11 representative task chains deployed on a self-driving bus with TensorRT~\cite{nvidia_tensorrt} and ROS2~\cite{ros2}. 
        Our results show that our schemes effectively prevent deadline misses, achieving a $61\%$ reduction in the deadline miss ratio and reducing kernel collisions by up to 51\% compared to state-of-the-art methods that require source code modifications. 

        \vspace{-0pt}\item Our proposed kernel scheduling strategies can be easily extended to similar CPS applications with predictable task graphs and dynamic data. Moreover, our investigation into GPU kernel launch mechanisms, such as overlapping batched launches and handling kernel collisions, provides rich insights to optimize GPU scheduling across various applications.
        \end{enumerate}

\vspace{-6pt}
\section{Preliminaries and Challenges} 
    
 \vspace{-1pt}\noindent\textbf{Autonomous Driving Application Characteristics.} 
    We use Autoware~\cite{kato2018autoware}, a widely adopted autonomous driving framework, to illustrate application characteristics. 
    Autoware implements a \textit{publish-subscribe} paradigm in ROS2~\cite{ros2}, allowing developers to create custom applications by connecting input/output topics. 
    While task chain topologies are known, individual task implementations often remain hidden within proprietary libraries. 
    Though chains can be configured with deadlines to ensure quality of service, maintaining hard real-time guarantees is challenging in unpredictable environments. 
    Factors like unsynchronized sensors and varying processing latencies make deterministic scheduling difficult.
    Given these uncertainties, many systems advocate soft real-time objectives, such as minimizing deadline miss rates~\cite{li2023red} rather than enforcing strict guarantees across all tasks.

    \label{sec:challenge}
    \begin{figure}[!t]
    \centering 
      \includegraphics[width=0.95\linewidth]{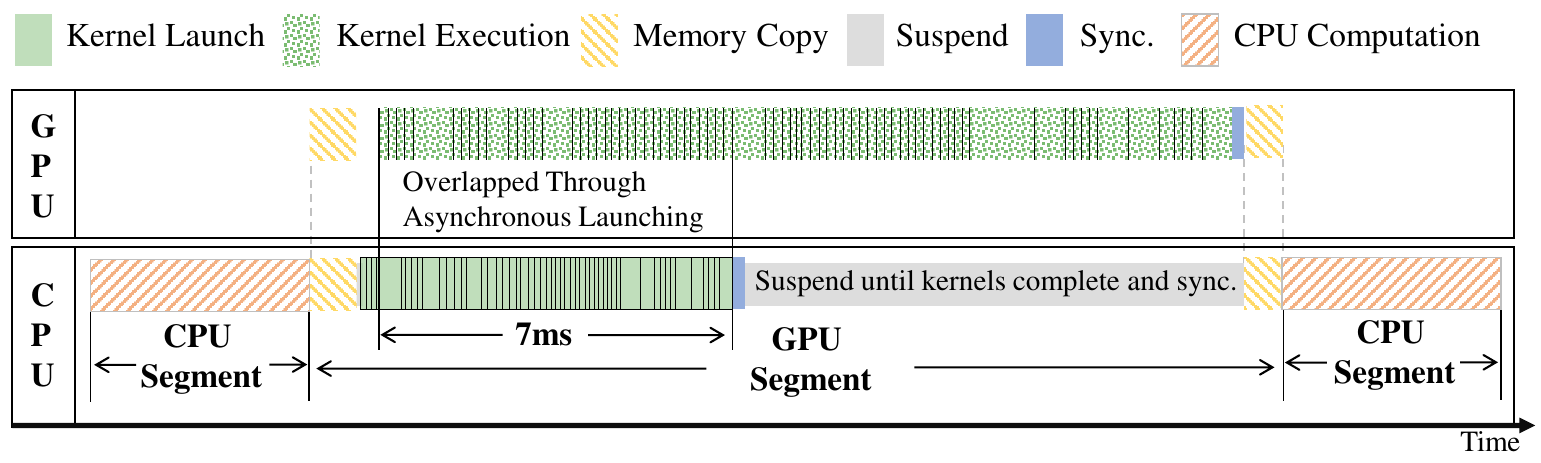} 
      \vspace{-12pt} 
        \caption{ \small The kernel launching and segmented structure for 2D object detection~\cite{ge2021yolox} using TensorRT~\cite{nvidia_tensorrt}.} 
        \label{fig:launch_vis}
        \vspace{-16pt}
    \end{figure}
    
 \vspace{2pt}\noindent\textbf{GPU Execution Patterns.} 
    Beyond application-level characteristics, GPU execution patterns significantly impact overall delay. 
    In CUDA GPUs, a stream orchestrates sequentially executed kernels, with
    each task's kernels bound to a single stream to preserve data dependencies.
    Kernels in each stream are launched from CPU to GPU, with optional synchronization points. Such sequential kernel launches create substantial CPU overhead. 
    As shown in Fig.~\ref{fig:launch_vis}, launching 323 kernels for a 2D object detection task takes 7~ms, accounting for 35\% of its 20~ms total GPU execution time.  
    As a result, commercial inference engines like TensorRT~\cite{nvidia_tensorrt} often adopt \emph{asynchronous launching} to minimize this overhead, inserting a single blocking synchronization function after the last kernel launch that suspends the CPU, as illustrated in Fig.~\ref{fig:launch_vis}, 
    until all kernels in the stream complete execution.  
 
 \vspace{2pt}\noindent\textbf{Kernel Launch Manipulation.}
    The default CUDA scheduler in closed-source drivers lacks awareness of task urgency and priority set by the CPU, leading to misaligned GPU executions. 
    To address this, we intend to modify how kernels are launched to better align GPU execution with CPU priorities. 
    The standard CUDA kernel launch API is shown as follows:  
    
    \vspace{1mm}
    \noindent\fbox{%
    \small
    \parbox{0.44\textwidth}{%
     cuLaunchKernel ($f_{p}$, $D_{g}$, $D_{b}$, $S_{m}$,  stream, args).
    }}\vspace{1mm}
    Here, $f_{p}$ is the pointer to the kernel function; $D_{g}$/$D_{b}$ denote grid/block dimensions; $S_{m}$ specifies dynamic shared memory; \emph{stream} determines the execution stream, and \emph{args} points to input/output data.  
    Of these, \emph{stream} is the only controllable parameter, offering preset priority levels (e.g., -5-0 for NVIDIA 3070Ti). Besides these parameters, we can also control kernel launch timing, e.g.,  delaying a kernel launch call.
    By carefully controlling these two factors, we aim to align GPU execution with the task's \emph{urgency level}.

\vspace{2pt}\noindent\textbf{Urgency Level.}
\label{subsection:challenges_scheduling}  
     We define a task chain's urgency level $UL_{C}(t)$, focusing only on GPU execution, as: \vspace{-4pt}
     \begin{equation} 
      UL_{C}(t) =  \dfrac{1}{ t_{arr} + D - \sum_{k=\Tilde{I^{gpu}}}^{N-1} {\Tilde{E^{gpu}_k}} - t},    
    \vspace{-3pt}
    \label{equal:urgency}
    \end{equation}
 where $t_{arr}$ is the arrival time of the task chain's current data frame; $t$ is the current time; $D$ is the task chain's deadline; $N$ is the total number of kernels in the task chain; $\Tilde{I^{gpu}}$ is the index of the currently executing kernel, and $\Tilde{E^{gpu}_k}$ is the estimated execution time for each kernel. 
 Since the kernels in the task chain are launched sequentially on the CUDA stream, the task chain's overall urgency is driven by the urgency of the currently active task/kernel.
    \begin{figure}[!t]
    \centering
        \begin{minipage}[t]{0.49\linewidth}
           \centering
            \includegraphics[width=\linewidth]{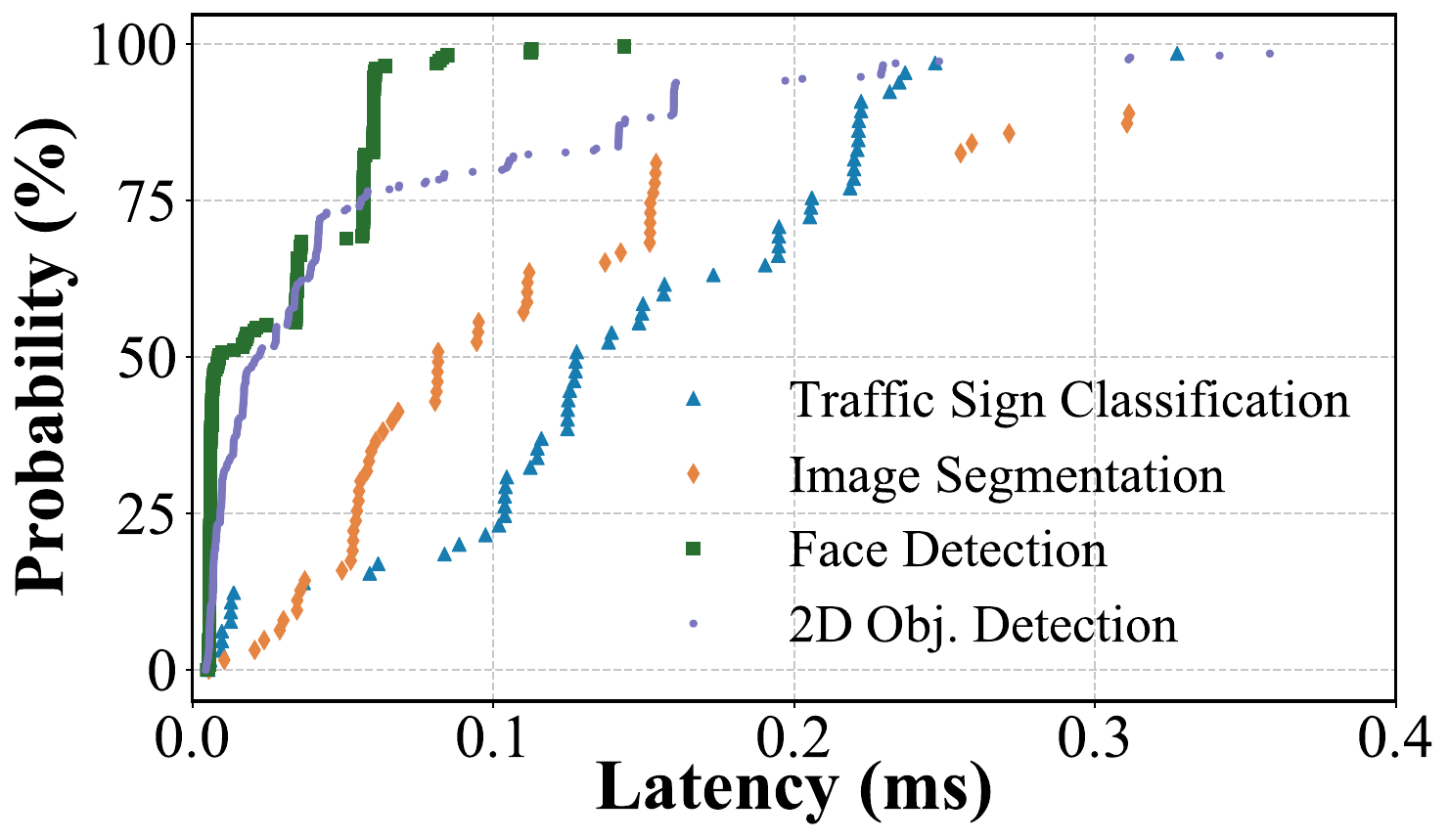}\\
            \captionsetup{width=0.95\linewidth}   \vspace{-12pt}
            \caption{ \small CDF of individual kernel execution time measured in NVIDIA 3070Ti.  } \label{fig:14_kernel_time} 
        \end{minipage} 
        \hfill
            \begin{minipage}[t]{0.49\linewidth}
            \centering 
            \includegraphics[width=\linewidth]{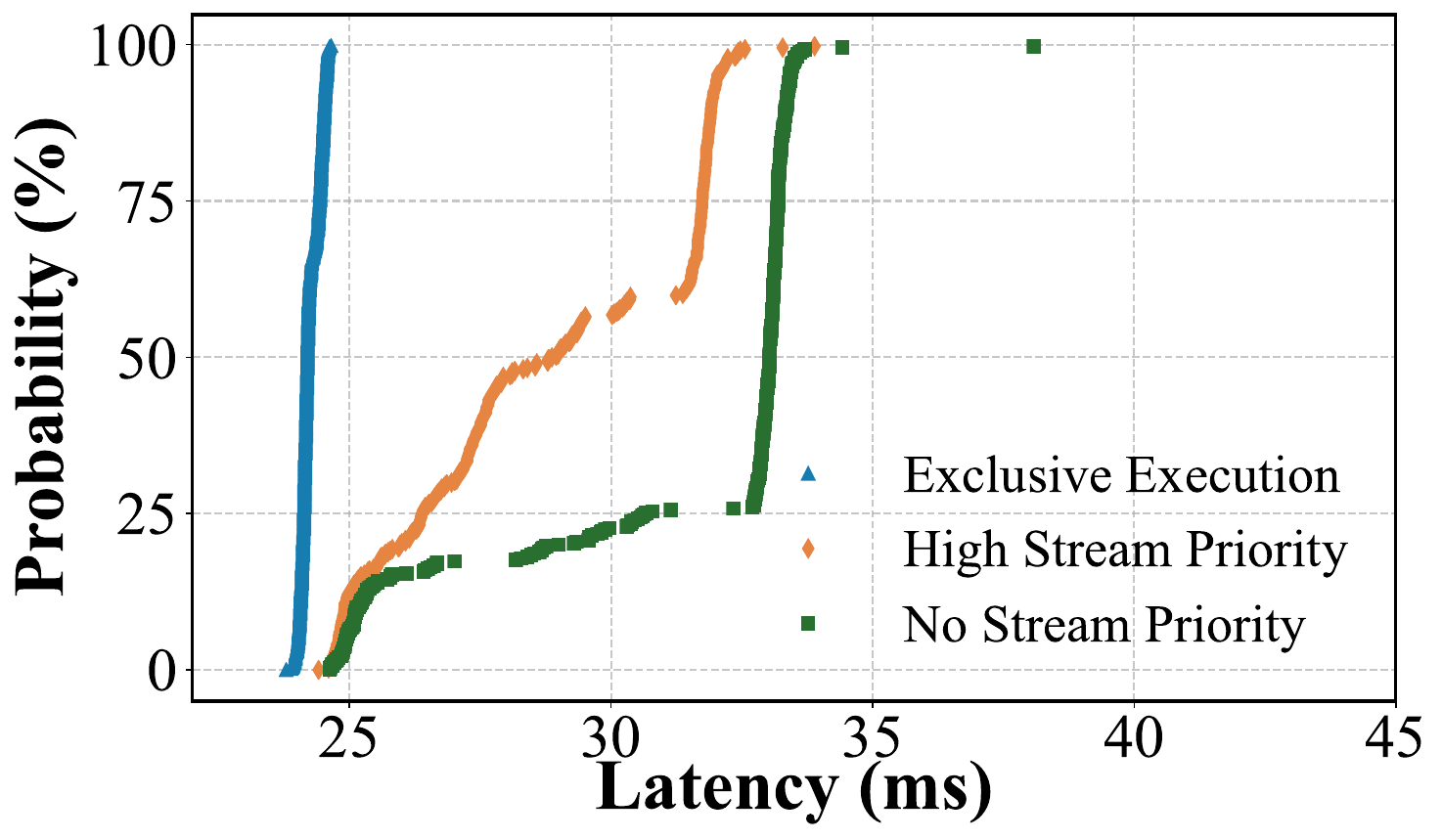}\\
            \captionsetup{width=0.95\linewidth}   \vspace{-12pt}
            \caption{ \small CDF of 2D detection latency when executed alongside 3D detection. }    
            \label{fig:13_cdf}  
        \end{minipage}
         \vspace{-10pt}
    \end{figure}

     \begin{figure*}[t]
    \centering
    \includegraphics[width=0.95\linewidth]{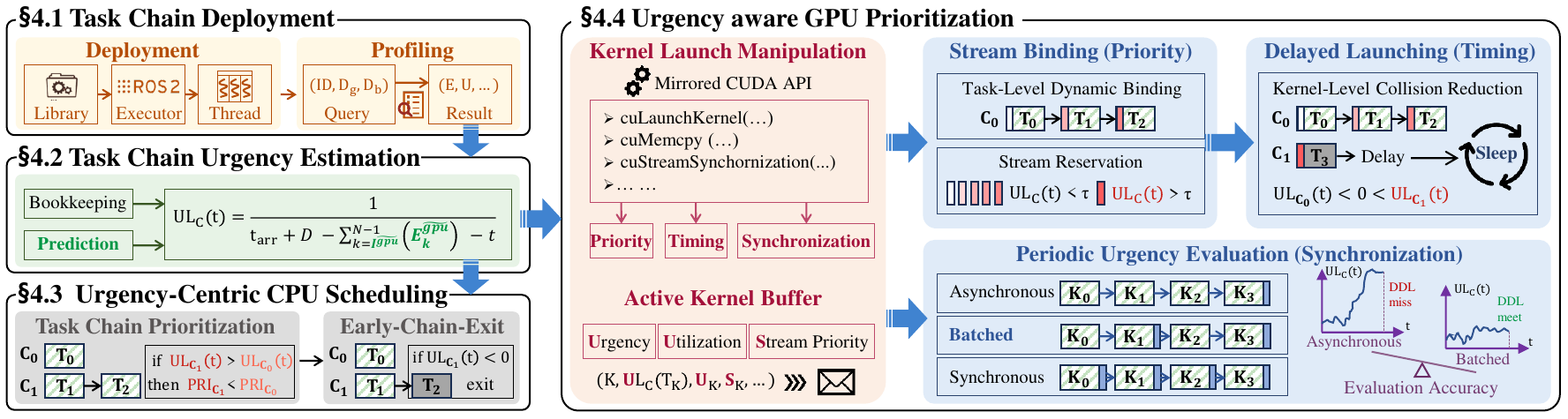} \vspace{-12pt}
    \caption{ \small System overview. \systemname provides refined control for both CPU/GPU prioritization within a unified framework.   
    } 
    \label{overview}    \vspace{-12pt} 
    \end{figure*}
    
\vspace{2pt}\noindent\textbf{Mismatch between Static Stream Priority and Dynamic Kernel Urgency Level.} \label{sec:mismatch} 
     We aim to schedule more urgent tasks earlier to decrease deadline misses. However, accurately evaluating a task's urgency level is difficult due to the challenge of tracking its remaining execution time, $\sum_{k=\Tilde{I^{gpu}}}^{N-1} {\Tilde{E^{gpu}_k}}$. 
     Accurately estimating the currently executing kernel $\Tilde{I^{gpu}}$ is feasible only with synchronous kernel launches. However, this approach incurs high launch and synchronization overhead, costing tens of microseconds -- comparable to the kernel execution time (e.g., under 100 \textmu s, as shown in Fig.~\ref{fig:14_kernel_time}). 
     When using asynchronous launches, we know which kernel is being launched,  but not which kernel is executed on the GPU, leading to inaccurate urgency level evaluation. 

    Furthermore, even with an accurate urgency level, prioritizing kernel executions remains challenging, largely due to the closed-source GPU drivers.
    While we can bind a task's kernels to a CUDA stream based on urgency, the limited number of priority levels (e.g., 6 in the NVIDIA 3070Ti) often leads to collisions among tasks.  
    Moreover, CUDA stream prioritization does not strictly enforce kernel execution order:
    resource contention~\cite{zhao2023miriam,li2023rosgm}, such as cache and memory bus contention, can still cause delays; 
    the non-preemptive nature of kernel block execution~\cite{pitfall,tx2} further exacerbates this issue, where low-priority kernels can block high-priority ones, resulting in missed deadlines. 
    For example, as shown in Fig.~\ref{fig:13_cdf}, sharing the GPU with 3D object detection~\cite{lang2019pointpillars} increases the end-to-end latency of 2D object detection~\cite{ge2021yolox} by 30\% at $95 percentile$, even when assigned a higher priority.

\vspace{-6pt}
\section{System Overview }  
 
\systemname transparently prioritizes task execution at runtime on both CPU and GPU based on each task's continuously changing urgency. 
Unlike existing solutions that either require source code access or rely on static task-to-stream priorities, \systemname introduces a holistic kernel launch manipulation framework integrated into our custom CUDA driver library.  
As illustrated in Fig.~\ref{overview}, this framework dynamically adjusts a task's CUDA stream priority (binding urgent tasks to higher-priority streams), launch timing (delaying less urgent kernels to prioritize critical ones), and synchronization mode (enabling efficient periodic urgency evaluation).
What distinguishes \systemname from state-of-the-art methods~\cite{enright2024paam,li2023rosgm,xiang2019pipelined,stream_pri} is its unified approach to managing kernel launches, addressing dynamic urgency needs during asynchronous and time-shared GPU operations.  

\noindent\textbf{Urgency Evaluation and CPU Prioritization}. 
\systemname leverages OS scheduling to implement urgency-aware CPU execution by setting chain-level priority based on each task chain's urgency level.
After deploying task chains to middleware entities, we profile kernel parameters including kernel execution times. At runtime, key variables like segment/kernel indices are estimated. 
These elements are combined to evaluate each chain's urgency, which is then mapped to appropriate CPU scheduler priorities.

\noindent\textbf{GPU Prioritization}.  
\systemname addresses the more complex challenge of GPU prioritization, due to the closed-source nature of the GPU libraries and drivers, by intercepting kernel launch APIs that overhaul the default CUDA dynamic library with a custom variant when CUDA function calls are invoked from applications. 
This interception enables manipulation of otherwise inaccessible parameters: \textit{stream priority}, \textit{launch timing}, and \textit{synchronization mode}. 
Specifically, \systemname addresses the following key design questions:  
 
\emph{1) To which priority-level CUDA stream do we bind a task's kernels}?  
By default, all kernels within a task are statically bound to a single CUDA stream, respecting data dependencies but remaining oblivious to urgency. 
\systemname introduces a dynamic, urgency-aware task-to-stream binding mechanism, allowing the stream priority to change across different instances (the arrival of the input data activates a new instance of a task ) without compromising execution coherence.
This flexibility, in contrast to static binding methods~\cite{xiang2019pipelined,enright2024paam},
enables dynamic prioritization based on task urgency. 
Additionally, to address the limited number of CUDA stream priority levels and the potential for kernel collisions, we devise a reservation scheme that reserves the highest stream priority for the most urgent task.

 \emph{2) How do we control the launch timing of kernels for tasks with varying urgency levels?}   
A critical challenge is the mismatch between dynamic task urgency and fixed stream priorities.    
For example, a task that has recently missed its deadline may retain a high-priority stream despite decreased urgency, while an increasingly urgent task might remain stuck with a low-priority stream allocation.  
\systemname resolves this through a \emph{delayed launching} mechanism inspired by wireless MAC protocols~\cite{ye2004medium}.  
When detecting potential kernel collisions between low-urgency and high-urgency kernels, \systemname briefly delays the low-urgency launch, making critical tasks execute first regardless of their stream assignment.
 
\emph{3) How frequently should we evaluate a task's urgency level}? 
While task urgency continuously changes as kernels launch and execute, frequent evaluations introduce costly synchronization overhead by serializing CPU launch/synchronization and GPU execution times. 
\systemname balances accuracy and efficiency through periodic urgency evaluation using \emph{batched kernel synchronization}, evaluating urgency after launching multiple kernels rather than individually.
We further optimize performance with \emph{batch overlapping}, allowing CPU kernel launches to proceed concurrently with GPU execution across batches.  

\vspace{-6pt}
\section{System Design}\label{sec:design}
This section presents the detailed design of \systemname, outlining our design goals, and diving into its core modules. 
\systemname supports transparent and efficient real-time scheduling for multi-task chains, with the following objectives:

\noindent\textbf{Deadline Compliance}. Minimize missed deadlines of task chains of time-sensitive autonomous driving applications; 
 
\noindent\textbf{Transparency}. Operate seamlessly with closed-source applications and GPUs without requiring access to source code.
 
\noindent\textbf{Efficiency}. Ensure low complexity and minimal delays to maintain fast, responsive GPU scheduling.
  
\vspace{-4pt}
\subsection{Task Chain Deployment}  
\begin{figure}[!t]
\centering
\includegraphics[width=0.9\linewidth]{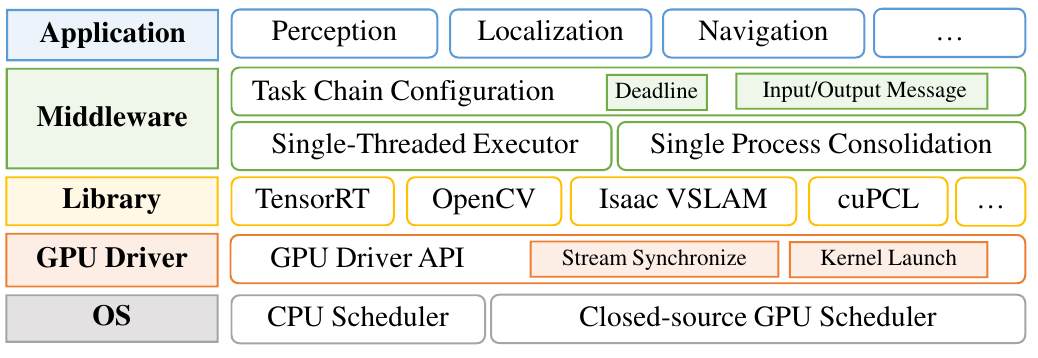} \vspace{-10pt}
\caption{ \small Components in an autonomous driving system.}
\label{fig:components} \vspace{-16pt}
\end{figure}

Autonomous driving systems are typically organized as modular task chains deployed using middleware frameworks like ROS2.  
The deployment process involves four key steps, following the approach in~\cite{kato2018autoware}: 
(1) implementing tasks using ROS2 interface with closed-source libraries like cuPCL~\cite{lidar_ai} for point cloud filtering, TensorRT~\cite{nvidia_tensorrt} for neural networks; 
(2) defining inter-task message formats; 
(3) connecting tasks into chains with application-defined deadlines; and
(4) mapping each task to a single-threaded ROS2 executor. 
To allow concurrent GPU execution of kernels from multiple executors within the same address space, we consolidate all executors into a single process. This setup ensures that \systemname operates cohesively and efficiently. 
Fig.~\ref{fig:components} shows the resulting system architecture after deployment.  
  
At this stage, we have multiple task chains deployed, each consisting of several tasks with CPU/GPU segments as illustrated in Fig.~\ref{fig:launch_vis}. 
We further propose incorporating a \emph{non-intrusive offline profiling stage}, where each chain processes test data in isolation and execution statistics are collected through API interception.  
This profiling captures the number of GPU kernels $N$ and CPU segments $M$ per chain, plus creates a lookup table (Tab.~\ref{tab:lookup}) recording key parameters: grid dimension $D_{g}$, block dimension $D_{b}$, execution time $E^{gpu}_k$, GPU utilization $U_k$, and the corresponding GPU segment ID. 
This lookup table maps input parameters to their corresponding kernel execution time and utilization, accommodating variations due to dynamic scene complexity.

 \begin{table}[b] \vspace{-12pt}
  \setlength\tabcolsep{2pt} 
 \renewcommand\arraystretch{1}
   \centering \scriptsize
\begin{tabular}{|c|c|c|c|c|c|c| }
\hline Kernel ID  &Grid Size $D_{g}$  & Block Size $D_{b}$   &  $E^{gpu}_k$ (ms)    & $U_k$(\%)     & GPU Segment ID  \\
\hline $0$ &22      & 512          &  0.028     & 19    & 0 \\
\hline $0$ &32      & 512          &  0.037     & 33    & 0 \\
\hline $1$ &40      & 512          &  0.044     & 42    & 1 \\
\hline  
\end{tabular} \vspace{-0pt}
 \caption{  \small  Example lookup table for path finder~\cite{rodinia}. 
 } \vspace{-2pt}
   \label{tab:lookup}
\end{table}
 
\vspace{-4pt}
\subsection{Task Chain Urgency Estimation}    
\label{sec:calculation}
    
 With task chains set up, \systemname is ready to operate at runtime. To facilitate the prioritization of task chain instances (i.e., each data frame initiates an instance of the task chain), we refine the definition of urgency level given in Eq.~\eqref{equal:urgency} to: \vspace{-3pt}
    \begin{equation} 
      UL_{C}(t) =  \dfrac{1}{t_{arr} + D - \sum_{k=\Tilde{I^{gpu}}}^{N-1} {\Tilde{E^{gpu}_k}} - \sum_{j=\Tilde{I^{cpu}}}^{M-1} {\Tilde{E^{cpu}_j}} - t},    
    \vspace{-2pt}
    \label{equal:urgency2}
    \end{equation}
where $UL_{C}(t)$ denotes the urgency level for task chain $C$, $t_{arr}$ denotes the arrival time of the input data frame, $D$ denotes the task chain's deadline which is usually made known by the applications, $N$ denotes the total number of GPU kernels in the task chain, $\Tilde{I^{gpu}}$ denotes the index of the currently executing kernel, and $\Tilde{E^{gpu}_k}$ denotes the estimated execution time for each kernel. Additionally, $M$ is the number of CPU segments in the task chain, $\Tilde{I^{cpu}}$ is the index of the currently executing CPU segment, $\Tilde{E^{cpu}_j}$ is the estimated execution time for each CPU segment, and $t$ denotes the current time.

The value of $UL_{C}(t)$ is evaluated in the runtime.  
On the one hand, we perform simple bookkeeping, such as the arrival time of the input data frame $t_{arr}$. 
On the other hand, we perform prediction of several important variables such as execution times of GPU kernels $\Tilde{E^{gpu}_k}$ and CPU segments $\Tilde{E^{cpu}_j}$, as well as the index of the currently executing GPU kernel $\Tilde{I^{gpu}}$ and CPU segment $\Tilde{I^{cpu}}$.
 
To predict the execution time of CPU segments $\Tilde{E^{cpu}_j}$ for the current data frame, we use the POSIX API \textit{clock\_gettime()} to record and calculate moving averages across recent instances as $\Tilde{E^{cpu}_j}$.  
For GPU kernels, instead of directly measuring potentially inaccurate  execution times due to GPU sharing and closed-source GPU drivers, we record grid size $D_{g}$ and block size $D_{b}$ for each kernel, then query the lookup table illustrated in Tab.~\ref{tab:lookup} to obtain the corresponding 
execution times. 
The averaging recent results are then to predict $\Tilde{E^{gpu}_k}$.   
 
Tracking the currently executing kernel index $\Tilde{I^{gpu}}$ is the most challenging. 
We maintain a kernel launch counter for each chain, reset at new instances and incremented by one with each kernel launch within the chain.
However, due to asynchronous kernel launching,  
$\Tilde{I^{gpu}}$ only reflects which kernel is being launched, not which is \emph{currently executing} on the GPU.
This ``kernel execute-launch gap" can lead to inaccurate urgency estimation, an issue addressed in Sec.~\ref{sec:periodic}. 
Once the last kernel of a GPU segment launches (identified during profiling), a new CPU segment starts with $\Tilde{I^{cpu}}$ incremented by 1.  
Finally, we evaluate chain urgency $UL_{C}(t)$ at three key points: when a new data frame arrives, when a new CPU segment starts, and when a new kernel launches. Tab.~\ref{chain_property} shows example results, including their total execution times.

 \begin{table}[b]  \vspace{-8pt}
\setlength\tabcolsep{1.8pt} 
 \renewcommand\arraystretch{1}
  \scriptsize
   \centering 
\begin{tabular}{|c|c|c|c|c|c|c|c|c|c| }
\hline    Chain  &  \thead{ \scriptsize{Modality}}     &  \thead{ \scriptsize{Period}  \vspace{-2pt}\\ \scriptsize{(ms)}}     &  \thead{ \scriptsize{ $D$ }  \vspace{-2pt}\\ \scriptsize{(ms)}}      & $E^{cpu}_{C}$ (ms)   & $E^{gpu}_C$ (ms) &  \thead{ \scriptsize{Urgency }  \vspace{-2pt}\\ \scriptsize{$UL_{C}(t)$ }}   &  \thead{ \scriptsize{Priority}  \vspace{-2pt}\\ \scriptsize{$PRI_C$} }     \\
\hline   $C_0$ &LiDAR    & 150  & 120     &  17.4 ($\pm$ 4.9) & 28.4 ($\pm$ 3.0)  &  0.0134  &  6      \\
\hline   $C_1$ &LiDAR    & 150  & 120     &  16.2 ($\pm$ 3.2) & 28.4 ($\pm$ 3.1)  &  0.0132  &  7      \\
\hline   $C_2$ &Camera    & 500 & 120      & 21.0 ($\pm$ 4.6)  & 27.0 ($\pm$ 1.3)  &  0.0138  &  5     \\
\hline   $C_3$ &Camera    & 200  & 120    & 20.2 ($\pm$ 1.7)  & 30.2 ($\pm$ 1.3)  &   0.0143  &  2      \\ 
\hline   $C_4$ &Camera    & 150 & 120      & 21.8 ($\pm$ 2.7)  & 19.5 ($\pm$ 2.8)  &  0.0127 &  8      \\ 
\hline   $C_5$ &Camera    & 200 & 120      &  20.2 ($\pm$ 1.7)  & 30.2 ($\pm$ 1.3)  & 0.0143  &  3       \\ 
\hline   $C_6$ &Camera    & 200 & 120      & 21.8 ($\pm$ 2.7)  & 19.5 ($\pm$ 2.8)  &  0.0127  &  9     \\ 
\hline   $C_7$ &Camera   & 500 & 120      &   21.0 ($\pm$ 4.6)  & 27.0 ($\pm$ 1.3)  &  0.0138  &  4        \\ 
\hline    {$C_8$} & LiDAR   & 200  & 120     & 21.3 ($\pm$ 3.9)  & 19.7 ($\pm$ 2.9)  & 0.0126  &  10     \\ 
\hline    {$C_9$} &Camera+LiDAR   & 500  & 120     &  11.2 ($\pm$ 1.4)  & 46.1 ($\pm$ 4.2)  & 0.0159  &  1      \\ 
\hline    {$C_{10}$} &Text   & 5000  & 200     &  17.8 ($\pm$ 4.6)  & 6.7 ($\pm$ 2.9)  &  0.0050  & 11      \\ 
\hline
\end{tabular} 
\vspace{0pt}
\caption{\small Characteristics of an example workflow. Please refer to Sec.~\ref{sec:implementation} for a detailed setting for $C_0$ to $C_{10}$. }
 \label{chain_property} \vspace{-4pt}
\end{table}

 \begin{table}[b]  \vspace{-12pt}
\setlength\tabcolsep{2.5pt} 
 \renewcommand\arraystretch{1}
  \scriptsize
   \centering 
\begin{tabular}{|c|c|c| }
\hline    Function Name        &  \thead{ \scriptsize{Timing}}   &  \thead{ \scriptsize{CUDA Stream Pirority}}      \\
\hline   \textit{cuLaunchKernel()}      &Delayed Launching   & Dynamic Stream Binding        \\
\hline   \textit{cuMemCpy()}            &Delayed Launching    &  No Stream Priority     \\
\hline   \textit{cuStreamSycnhronize()}   &  Zero Load    & Consistent with \textit{cuLaunchKernel()}     \\
\hline
\end{tabular} 
\vspace{0pt}
\caption{\small   Intercepted function calls and their modifications.}
   \label{tab:funtion_call} \vspace{-3pt}
\end{table}

\vspace{-4pt}
\subsection{Urgency-Centric CPU Scheduling}  
    Each time a task chain starts a new CPU segment at $t_0$, we calculate $UL_{C}(t_0)$ and then map the urgency level to the corresponding CPU priority $PRI_{C}$ for all active chains. 
    We do so by ranking their urgency levels at $t_0$. 
    More urgent task chains receive a lower $PRI_{C}$, which translates to a higher CPU priority.  
    This priority $PRI_{C}$ is further normalized to $\bar{PRI_C} \in (1, NUM\_PRI)$, where $NUM\_PRI$ represents the total number of priority levels supported by the operating system. 
    Tab.~\ref{chain_property} illustrates the chain priority assignment using an example workflow.  
    This priority is then passed into the CPU scheduler to prioritize task threads:
    
    \noindent\fbox{%
    \small
    \parbox{0.44\textwidth}{%
      sched\_setscheduler(thread\_id, SCHED\_FIFO, priority=$\bar{PRI_C}$);
      }
      }
 
     In addition, \systemname implements an \emph{early-chain-exit} mechanism to conserve resources. Suppose a task starts execution at time $t_1$. If the chain urgency  $UL_{C}(t_1) < 0$, then we exit the chain without executing the remaining tasks.

\vspace{-4pt}
\subsection{Urgency-Aware GPU Prioritization through Transparent Kernel Launch Manipulation}
\subsubsection{Kernel Launch Manipulation through API Interception}
The core innovation of \systemname lies in prioritizing kernels that are launched and executed on GPU.  
The closed-source nature of third-party libraries makes it impossible to directly implement such prioritization in the library. 
Thus, we develop an API interception technique to achieve GPU prioritization through transparently manipulating kernel launch parameters such as priority, timing, and synchronization. 
  
\begin{figure}[!t]
    \centering
    \includegraphics[width=1.0\linewidth]{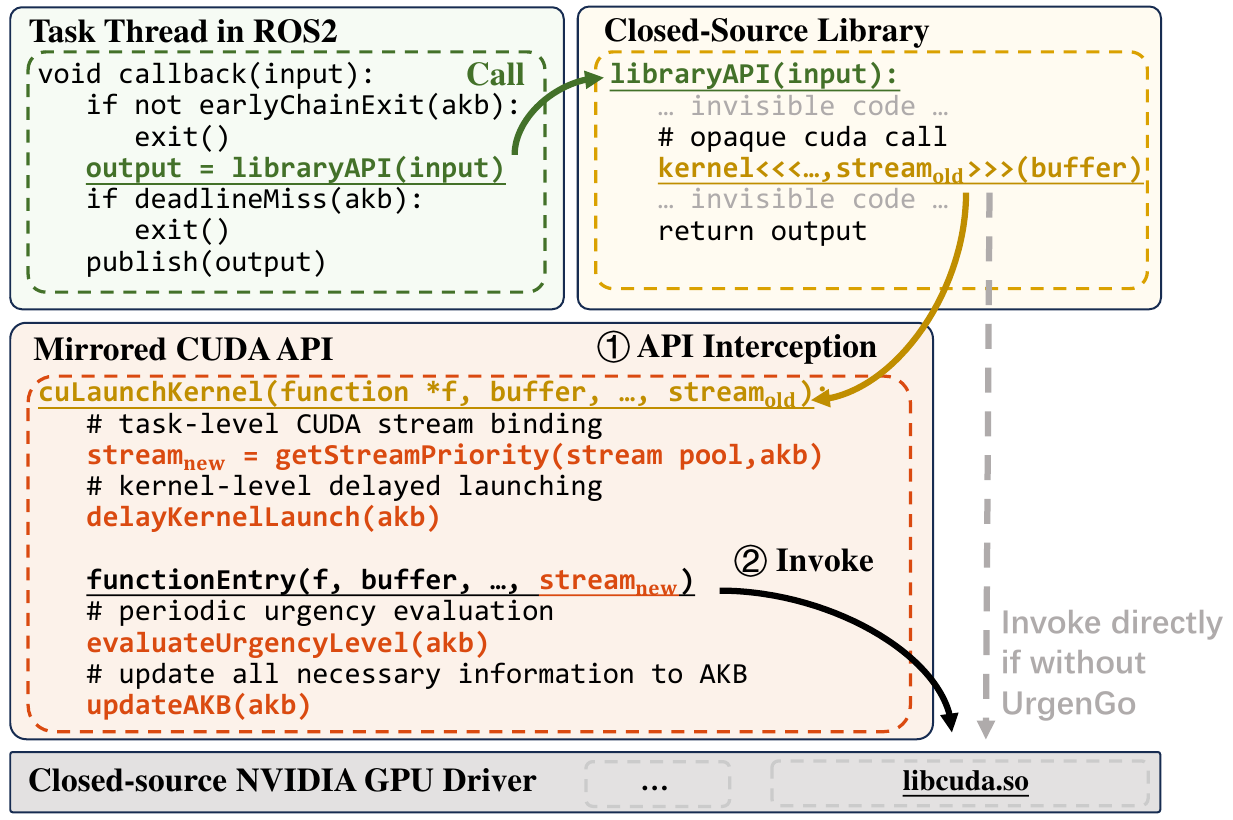} \vspace{-15pt}
    \caption{ \small Kernel manipulation with a mirrored CUDA API. 
    }
    \label{fig:stack} \vspace{-8pt}
    \end{figure}

In particular, we create a mirrored version of the CUDA driver library, using the POSIX API \textit{dlsym()} to wrap CUDA functions from the original driver library.
By modifying the environment parameters \textit{$LD\_LIBRARY\_PATH$}, we instruct the Linux dynamic linker \textit{ld.so} to load our custom mirrored library at runtime. 
When a task invokes CUDA function calls, such as \textit{cuLaunchKernel()} shown in Fig.~\ref{fig:stack} (see its API details in Sec.~\ref{sec:challenge}), the dynamic linker resolves the symbols and redirects the calls to our custom implementations. 
 
Here, we mainly intercept the function calls corresponding to the following three operations: kernel launch, memory copy, and stream synchronization, as summarized in Tab.~\ref{tab:funtion_call}.  
Among these three operations, we focus primarily on scheduling kernel launches because they offer the most prioritization capability by manipulating both the timing and the stream priority of the kernels.  Memory copy operations rank second by supporting the timing control. The last category, stream synchronization,   provides various synchronization mechanisms. When teamed with  \textit{cuLaunchKernel()}, kernel launch synchronization is achieved.
 
\subsubsection{Active Kernel Buffer}   
Once a kernel launch is intercepted, the kernel to be launched becomes ``active'' in our notion, until it completes execution and synchronizes with the \textit{cuStreamSynchronize()/cuEventSynchronize()}. To record the state of the active kernel, we define a data class $ActiveKernelBuffer$ (\buffername) in \textit{component\_manager.hpp} within ROS2, which is shared among all task chains.  
Each task chain maintains its own instance of \buffername to prevent race conditions. 
For each active kernel $K$, we maintain its \buffername entry $(K, U_{K}, S_K, C, \bar{PRI_C}, T_K,  UL_{C}(T_K))$, with
     \emph{Kernel ID} $K$,
     \emph{Profiled GPU utilization} $U_{K}$, 
     \emph{Stream ID} $S_K$ which denotes the ID of the CUDA stream this kernel is bound to,  
     \emph{Chain ID} $C$, 
     \emph{CPU priority} $\bar{PRI_C}$.
     \emph{Most recent urgency level evaluation timestamp} $T_K$, and the \emph{last-evaluated urgency level} $UL_{C}(T_K)$. 
Once the kernel completes execution and synchronizes, its record is deleted from the \buffername.

The overhead for accessing/updating \buffername is minimal, e.g., 0.5 \textmu s for updating with an 11th Gen Intel Core i7-11800H processor, and its memory consumption is also modest, amounting to 1KB for all 9 task chains described in Tab.~\ref{chain_property}.

  \begin{figure}[t]
  \centering 
   \subfloat{    \includegraphics[width=\linewidth]{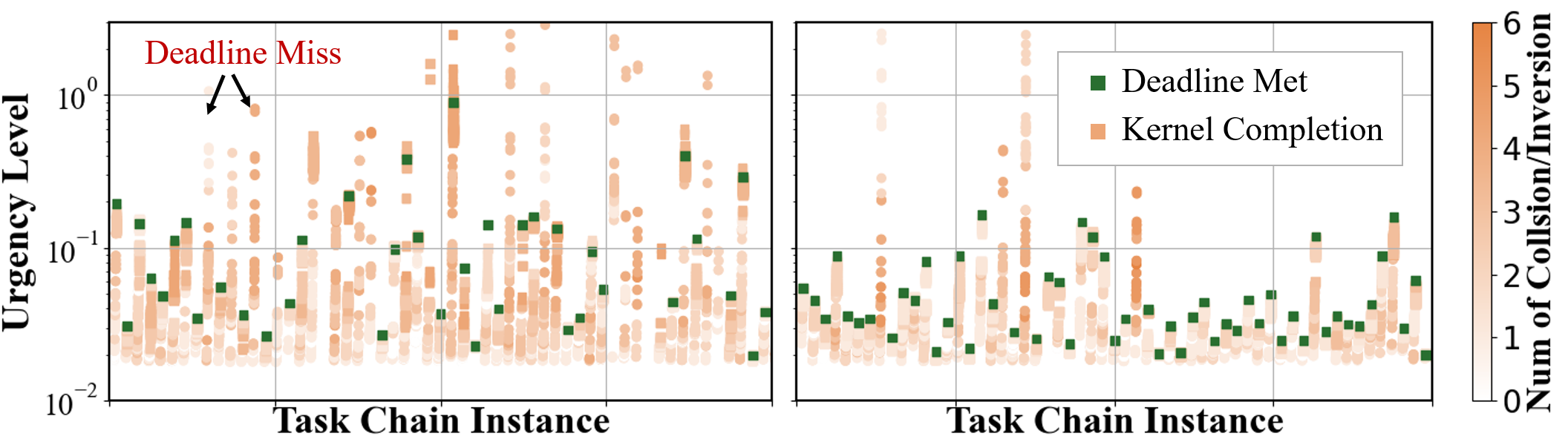}}\\
   \vspace{-1pt}
   \small  \hspace{0.0cm} (a) Without Stream Binding   \hspace{0.3cm}  \small   (b) With Stream Binding   \\
   \vspace{-10pt}
  \caption{ \small Estimated urgency and the number of collisions across different instances of $C_2$, when co-running $C_0$ to $C_7$. Compared to (a), (b) has many more lighter-shaded points which indicates significantly fewer collisions.   
  \label{fig:early_result}} 
\vspace{-16pt}
\end{figure}
 
\subsubsection{Task-Level Stream Binding with Reservation}  
When a task's first kernel is about to be launched, we intercept the kernel launch API and perform stream binding, which involves binding the kernel to a CUDA stream dynamically with a suitable priority level to ensure urgency-aware execution.  
Specifically, \systemname maintains a pool of CUDA streams for each task, with each stream assigned a different priority. The number of streams in this pool equals the maximum number of stream priorities ($NUM\_PRI$) supported by the GPU hardware. 
Upon intercepting a kernel launch,  \systemname replaces the original stream identifier $stream_{old}$,  with a new stream identifier $stream_{new}$, as illustrated in Fig.~\ref{fig:stack}. 
Once $stream_{new}$ is selected for the first kernel, all subsequent kernels in that task are bound to the same $stream_{new}$ to respect data dependencies between these kernels. 
 
However, selecting $stream_{new}$ for each task is non-trivial due to the limited number of available priorities ($NUM\_PRI$), a large number of tasks, as well as unpredictable launch times. It is likely to have the problem of \emph{priority collision} -- i.e., a very urgent task is bound to the same stream priority as a non-urgency task, or even worse \emph{inverted binding} -- i.e., a very urgent task is inversely bound to a stream with lower priority than a non-urgency task. We collectively refer to these issues as \emph{kernel collision}.
To address this, we propose to reserve the highest stream priority (i.e., -5) for those tasks whose urgency are above the \emph{truly urgent} threshold $TH_{urgent}$.  

To properly set up the threshold, we profile the urgency level distribution by periodically recording the highest urgency value among all active kernels in \buffername. 
We then calculate the 95th percentile urgency from the collected data to set $TH_{urgent}$.  
If a task's urgency level $UL_{C}(t)$ exceeds this threshold -- indicating a higher likelihood of missing its deadline -- it is assigned the highest stream priority. For the remaining tasks, we access \buffername and rank their $UL_{C}(t)$ values, then normalize these rankings to the range $(1, {NUM\_PRI}-1)$ to select $stream_{new}$ accordingly. 
With this binding policy, we effectively reduce the total number of kernel collisions, as illustrated in Fig.~\ref{fig:early_result}, where the presence of lighter-shaded points indicates fewer collisions.

   \begin{figure}[t]
  \centering 
   \subfloat{   \includegraphics[width=1.0\linewidth]{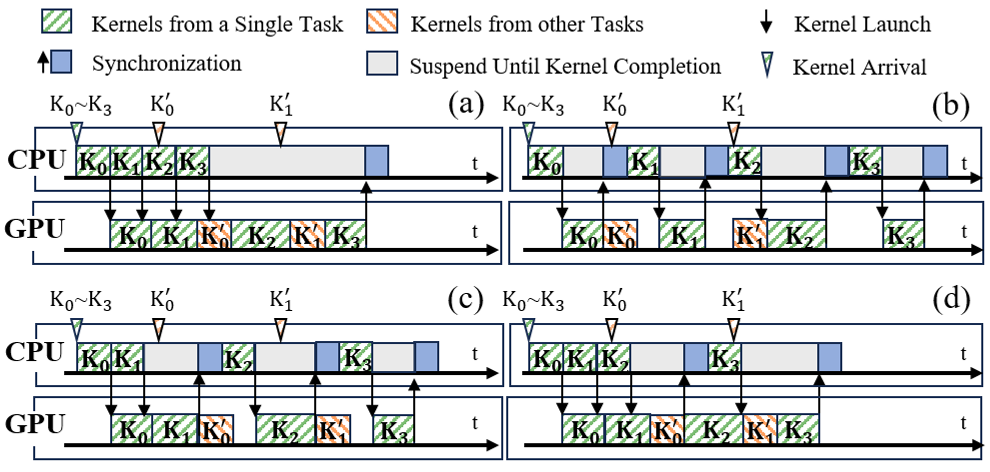}}\\
   \vspace{-9pt}
  \caption{ \small Urgency evaluation methods: (a) asynchronous launching renders urgency evaluation challenging, (b) synchronous launching with per-kernel evaluation, (c) synchronous batched launching, (d) overlapping batched launching.   \label{fig:launch}} 
\vspace{-16pt}
\end{figure}

\subsubsection{Delayed Kernel Launching}  
Since multiple task chains execute simultaneously, there are often multiple active kernels whose executions may collide on the GPU. To minimize deadline misses, it is important to prioritize these kernels and have the most urgent kernel execute early and if possible execute alone. Please note that the task-level stream binding only considers the urgency level when the first kernel of a task is about to be launched, without considering the dynamically changing urgency level as kernels of the same task proceed.  
For example, a task that has just missed its deadline will likely have a significant drop in urgency but may remain bound to a high-priority stream. 

To minimize collisions among active kernels, we learn from the rich experience in wireless MAC protocols~\cite{ye2004medium} and propose \emph{delayed launching}. When a task is about to launch a kernel $K$, it first accesses \buffername to check the urgency levels of all other active kernels. If there exists a kernel with an urgency level that exceeds the high urgency threshold $TH_{urgent}$, the task delays kernel $K$'s launch by entering a sleep loop.  
During the sleep loop, the task periodically sleeps and wakes up (e.g., every 1~ms used in our evaluation) to evaluate its urgency and consult \buffername. If the urgent kernels are still in \buffername, the task remains in the loop; otherwise, it exits the loop and proceeds to launch kernel $K$ and subsequent kernels in the task. Here, we choose a relatively large delay duration for two reasons: (1) the subsequent kernels for the truly urgent kernel are usually also truly urgent, and (2) a longer sleep can better amortize the context switch overhead. 
Finally, please note that kernels with extremely low GPU utilization (e.g., less than 0.1) are exempt from the delayed launching process, as they are considered less disruptive.
 
\subsubsection{Periodic Urgency Evaluation through Batched Kernel Launching}  
\label{sec:periodic}   
As mentioned earlier in Sec.~\ref{sec:calculation}, tracking which kernel is being executed on the GPU is challenging due to the lack of synchronization for kernel launches.  
Firstly, the default asynchronous kernel launching mode illustrated in~Fig.~\ref{fig:launch}(a), only requires kernel synchronization after launching the last kernel of a stream. As a result, we are unable to obtain the index of the kernel that is executing on the GPU at a given time. 
This will hinder the accurate estimation of the urgency level during the execution of a stream of kernels.
Secondly, a straightforward method to address this issue is to adopt synchronous kernel launching, which inserts synchronization calls \textit{cudaStreamSynchronize()} after each kernel launch to track kernel completion, as illustrated in Fig.~\ref{fig:launch}(b).
However, with such synchronous mode, the kernel launch/synchronization time on the CPU and the kernel execution time on the GPU are serialized. Given that both times can be substantial (e.g., accounting for 7~ms for launching 323 kernels in 2D object detection, and 10\textasciitilde200 \textmu s for each synchronization call in NVIDIA 3070Ti), this serial execution pattern incurs an excessive overhead.

To strike a balance between accurate kernel execution tracking and low launching overhead, we propose \emph{batched kernel launching}, where a synchronization call is inserted after launching a batch of kernels rather than after each kernel, enabling \emph{periodic urgency evaluation}
, illustrated in Fig.~\ref{fig:launch}(c). 
Specifically, we define a maximum interval $\Delta_{eval}$, during which kernels are launched without synchronization, as long as the sum of their estimated GPU execution time remains less than $\Delta_{eval}$. 
Once this sum exceeds the interval, a synchronization call is inserted, and a new batch is started.

Batched kernel launching improves overhead overlap, but synchronizes each batch. 
Small $\Delta_{eval}$ values may still cause synchronization bottlenecks.
To alleviate it, we employ a \textit{batch overlapping} scheme, where synchronization calls wait for the previous batch to complete, rather than the current one. 
We implement this using lightweight CUDA events to mark batch starts and \textit{cuEventSynchronize(CUevent event)} to wait for specific launching events.  
This allows \systemname to maximize the overlap between CPU kernel launching overheads and GPU executions, minimizing periodic urgency evaluation overhead.  
To ensure that all the overheads are fully overlapped, we set $\Delta_{eval}$ to a slightly larger value of $0.5~ms$, without compromising estimation accuracy.

\vspace{-8pt}
\section{ Implementation } 
 \label{sec:implementation} 
   We have implemented a prototype \systemname system in C++ with about 4.5K lines of code. The source code will be released before the conference. Below we list five key functions:

     \vspace{1mm}
     \noindent\fbox{%
     \small
     \parbox{0.46\textwidth}{ 
     \textcolor{codecolor}{createStreamPool}(Stream Pool) $\rightarrow$ void; \\ 
     \textcolor{codecolor}{getStreamPriority}(Stream Pool, \buffername) $\rightarrow$ stream ID; \\ 
      \textcolor{codecolor}{delayKernelLaunch}(\buffername) $\rightarrow$ flag;  \\
      \textcolor{codecolor}{evaluateUrgencyLevel}(\buffername) $\rightarrow$ urgency;\\
      \textcolor{codecolor}{updateAKB}(\buffername) $\rightarrow$ void;
     }}\vspace{1mm}

\noindent\textbf{Experiment Platform.}
        We implement \systemname using ROS2 Foxy Fitzroy~\cite{ros2} on Ubuntu 20.04.   
        The experiment platform features an industrial-grade Nuvo-8108GC~\cite{Nuvo-8108GC} server, designed for in-vehicle applications. It is equipped with an NVIDIA RTX 3070Ti GPU operating at 1.44 GHz, and an Intel(R) Core(TM) i7-9700E CPU processor (8 cores) running at 2.6 GHz. The GPU has 46 stream multiprocessors (SMs), each containing 128 CUDA cores and 4 tensor cores.  All GPU tasks are written and compiled with CUDA 11.4. 
 
\noindent\textbf{Task Chain Setup.}
{We collaborate with a self-driving start-up to develop a realistic autonomous navigation application, comprising 11 task chains across 6 service domains. As shown in Tab.~\ref{chain_property},
the 3D perception chains ($C_0, C_1$) integrate PointPillars~\cite{lang2019pointpillars} for 3D detection, and particle filtering~\cite{rodinia} for tracking;
the 2D perception chains ($C_2$ to $C_7$) utilize 2D object detection~\cite{ge2021yolox}, face detection~\cite{face}, traffic sign classification~\cite{resnet}, and segmentation~\cite{long2015fully}; 
the localization and navigation chain ($C_8$) employs ICP for LiDAR-based SLAM~\cite{duan2022pfilter}, feeding results into a path finding algorithm for global/local planning; 
the calibration chain~\cite{my_calibration} ($C_9$) leverages OpenCV and cuPCL~\cite{lidar_ai} for fast feature extraction and alignment;
the interaction chain ($C_{10}$) runs a LLAMA2-7B~\cite{touvron2023llama} for text generation. 
These chains execute periodically with a $15~ms$ jitter  
using closed-source frameworks like TensorRT~\cite{nvidia_tensorrt} and cuPCL~\cite{lidar_ai}. Profiling results (Tab.~\ref{profiling}) provide key performance metrics for each task, including kernel count $N$, GPU execution time $E_{gpu}$, and input size $N_s$. Input size varies based on the task -- image resolutions for 2D tasks, point cloud count for 3D detection and registration, particle count for tracking,  map size for planning, and text length for LLM processing. }

   \begin{figure}[!t]
    \centering
        \includegraphics[width=0.9\linewidth]{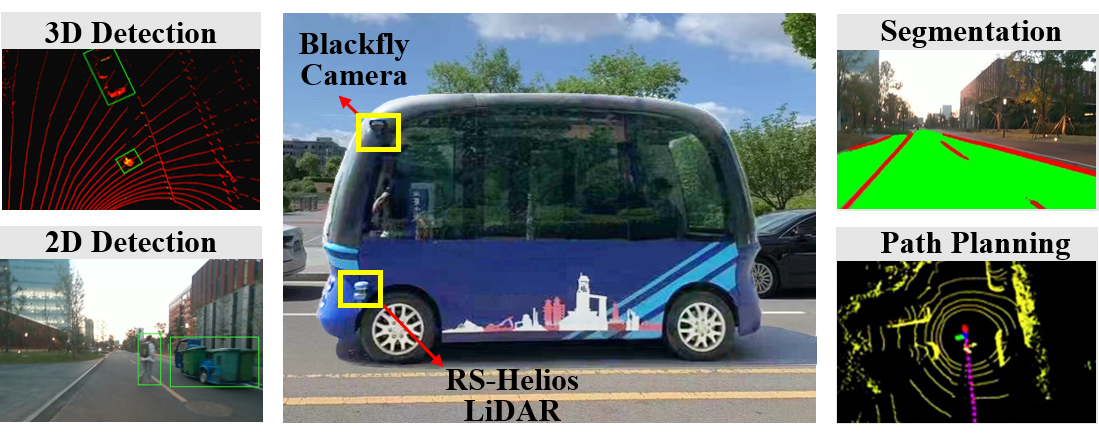}   
    \vspace{-14pt}
    \caption{ \small Data collection and evaluation are conducted on a self-driving bus, which is equipped with 3 LiDARs and 6 cameras. 
    The collected data is visualized to showcase the output from different task chains. 
    }\label{fig:car}  \vspace{-16pt}
\end{figure}

          \begin{table}[b] \vspace{-8pt}
          \setlength\tabcolsep{0.2pt} 
         \renewcommand\arraystretch{1}
           \centering \scriptsize
        \begin{tabular}{|c|c|c|c|c|c|c|  }
        \hline Task   & Input Size $N_s$   & \thead{ \scriptsize{\# of} \vspace{-2pt} \\ \scriptsize{Kernels $N_k$}} &TensorRT   & \thead{$E_{gpu}$ (ms) }     \\
        \hline  3D Object Detection~\cite{lang2019pointpillars}  &  22\,000$\leq N_s \leq$25\,000  & 41     & Y      & 13.4 ($\pm$ 1.3)        \\
        \hline  Particle Filtering~\cite{rodinia}    &  2\,500$\leq N_s \leq$8\,000    & 16   & N      & 15.0 ($\pm$ 2.8)       \\
        \hline  2D Object Detection~\cite{ge2021yolox}    &  $N_s$=1x640x640    & 323   & Y       & 19.8 ($\pm$ 1.2)        \\
        
        \hline  Face Detection~\cite{face}   &  $N_s$=10x112x112    & 225  & Y         & 7.1 ($\pm$ 1.3)     \\
        \hline  Traffic Sign Classification~\cite{resnet}    &  $N_s$=1x224x224    & 65    & Y         & 10.4 ($\pm$ 1.2)     \\
        \hline  Image Segmentation~\cite{long2015fully}   &  $N_s$=1x640x640     & 63    & Y         & 11.5 ($\pm$ 1.2)    \\
        \hline  Path Finding~\cite{rodinia}  & 6\,400$\leq N_s \leq$15\,360      & 256     & N       & 8.0 ($\pm$ 2.9)        \\
        \hline      {ICP Registration}~\cite{lidar_ai}  &  22\,000$\leq N_s \leq$65\,000    & 40     & N     & 21.3 ($\pm$ 3.9)        \\ 
        \hline       {Online Calibration}~\cite{my_calibration}  &   $N_s$=1x640x640     & 133     & N       & 11.2 ($\pm$ 1.4)      \\ 
        \hline     {LLAMA2-8B}~\cite{touvron2023llama}   &  $N_s \leq$  400     & 1106  & N    & 17.8 ($\pm$ 4.6)    \\
        \hline
        \end{tabular} \vspace{-2pt}
         \caption{GPU task profiling results.} \vspace{-2pt}
           \label{profiling}
        \end{table} 
 
\vspace{-6pt}
\section{Evaluation}

\subsection{Experiment Setting}   
To evaluate \systemname, we implemented an autonomous navigation application on a self-driving bus equipped with cutting-edge sensors: 3 Robosense RS-Helios LiDARs and 6 FLIR Blackfly color cameras. 
Our data collection spanned an extensive two-week period with ROSBAG, recording diverse scenarios across a 480,000 square meter testing campus, under various environmental conditions. 
The LiDAR point clouds were captured at 10FPS, while image streams were recorded at 15FPS.  
Our evaluation consisted of two phases: {(1) a \textit{trace-based} evaluation, where data was recorded and replayed using ROSBAG to isolate scheduling effects from navigation behavior, ensuring a repeatable environment for fair comparison. Each experiment lasted 10 minutes;} and (2) a \textit{long-term} evaluation, where all data was generated and processed in real-time over a two-hour period, providing deeper insights into system performance. 
Fig.~\ref{fig:car} shows our vehicle and sample data collected during experiments.

\vspace{-6pt}
\subsection{Evaluation Metrics}   
The primary metric used is the \textit{overall deadline miss ratio}, defined as the average of individual miss ratios across multiple task chains.
It is calculated as follows:
\vspace{-6pt}
\begin{equation}\small
    \text{Overall Deadline Miss Ratio} = \dfrac{1}{N} * \sum_{i=0}^{N-1} \dfrac{M^{miss}_{C_i}}{M^{total}_{C_i}},   \vspace{-6pt}
\end{equation}
where $M^{miss}_{C_i}$ and $M^{total}_{C_i}$ represent missed and total task instances of chain $C_i$, and $N$ is the total number of chains.  
The second metric is \textit{task chain latency}, reflecting the execution time and overhead introduced by \systemname. The execution time of each task chain is reported in Tab.~\ref{chain_property}. 
To ensure a comprehensive evaluation, we {divide the 11 task chains into two workflows: the first includes $C_0$ to $C_9$, and the second includes $C_6$ to $C_{10}$. We use a base deadline of 120~ms and set arrival rates as the reciprocal of the periods listed in the table. 
To simulate varying deadline constraints, we modify the base setting by halving the deadlines for a subset of task chains, controlled by a factor $f_{tight}$.
Additionally, we scale the base arrival rates by a factor $f_{a}$, ranging from [0.5, 1.3], and the base deadlines by a factor $f_{d}$, ranging from [0.7, 1.5]. This setup effectively simulates real-world conditions. 
Unless stated otherwise, we use task chains $C_0$ to $C_9$, set $f_{tight}$ to $40\%$, $f_{d}$ to 1.0, and $f_{a}$ to 1.0 as the default setting. }

 \vspace{-6pt}
\subsection{Comparison with Baseline Schedulers}
\noindent\textbf{Baseline Schedulers.} 
{
We compare \systemname against four baselines: PAAM~\cite{enright2024paam}, dCUDA~\cite{dcuda}, cCUDA~\cite{ccuda}, and vanilla CUDA. 
PAAM~\cite{enright2024paam}, the latest state-of-the-art scheduler for multi-task-chain applications, assigns predefined criticality levels to each chain and statically sets CPU and GPU priorities using the CAPA policy~\cite{picas}. Tasks with tighter deadlines are assigned higher criticality, determined via offline profiling. 
dCUDA is a dynamic GPU scheduling framework optimized for multi-tenant GPU serving. It maximizes GPU utilization by grouping kernels with complementary occupancy levels.  Within each group, low-utilization kernels receive higher CUDA priority, while groups execute in a Round-Robin manner. 
cCUDA introduces kernel splitting, a new dimension for kernel manipulation. It categorizes kernels into compute-bound or memory-bound,  then splits them into sub-kernels for optimized co-scheduling, improving concurrency and resource utilization.
}
 
\begin{figure*}[!t]
\centering
    \begin{minipage}[t]{0.25\linewidth}
        \centering 
        \includegraphics[width=0.9\linewidth]{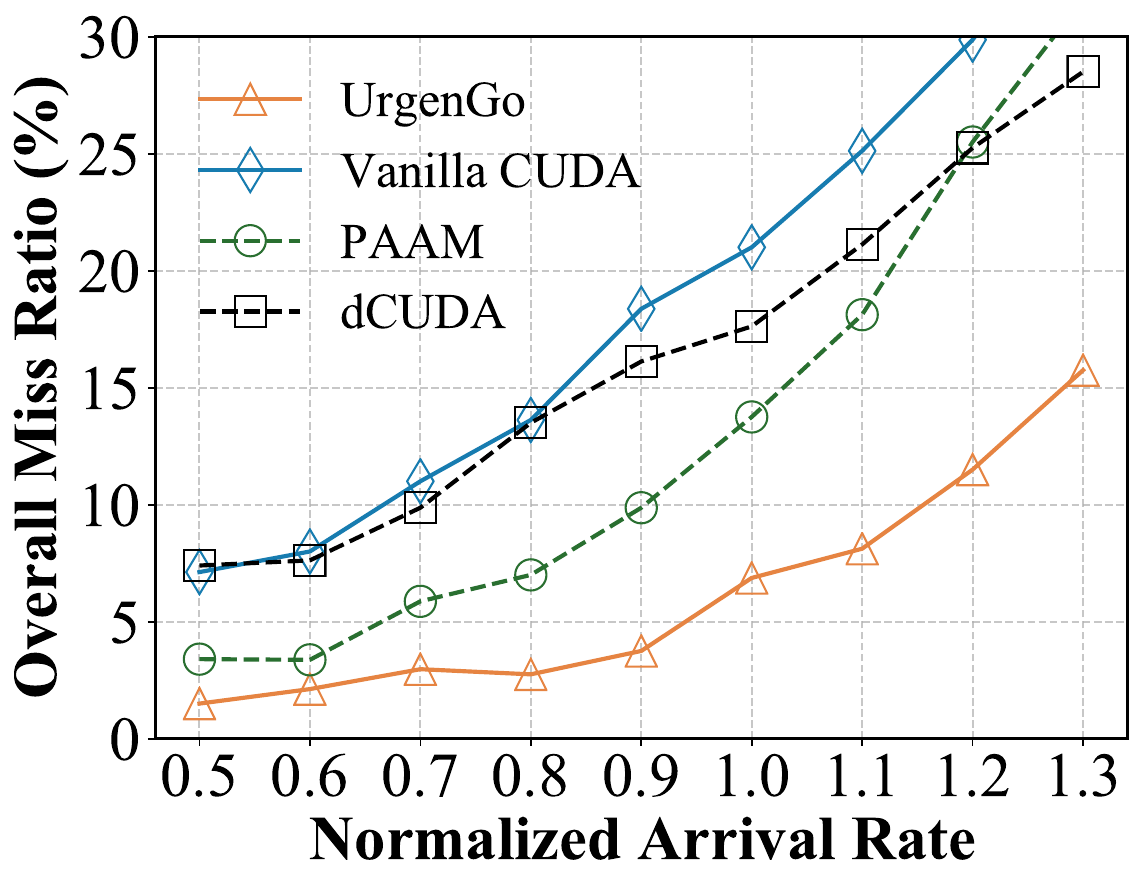}\\
    \captionsetup{width=0.95\linewidth} \vspace{-12pt}
    \caption{ \small {Performance with different arrival rates.}   } \label{fig:0_overall} 
    \end{minipage}%
    \begin{minipage}[t]{0.25\linewidth}
       \centering
        \includegraphics[width=0.9\linewidth]{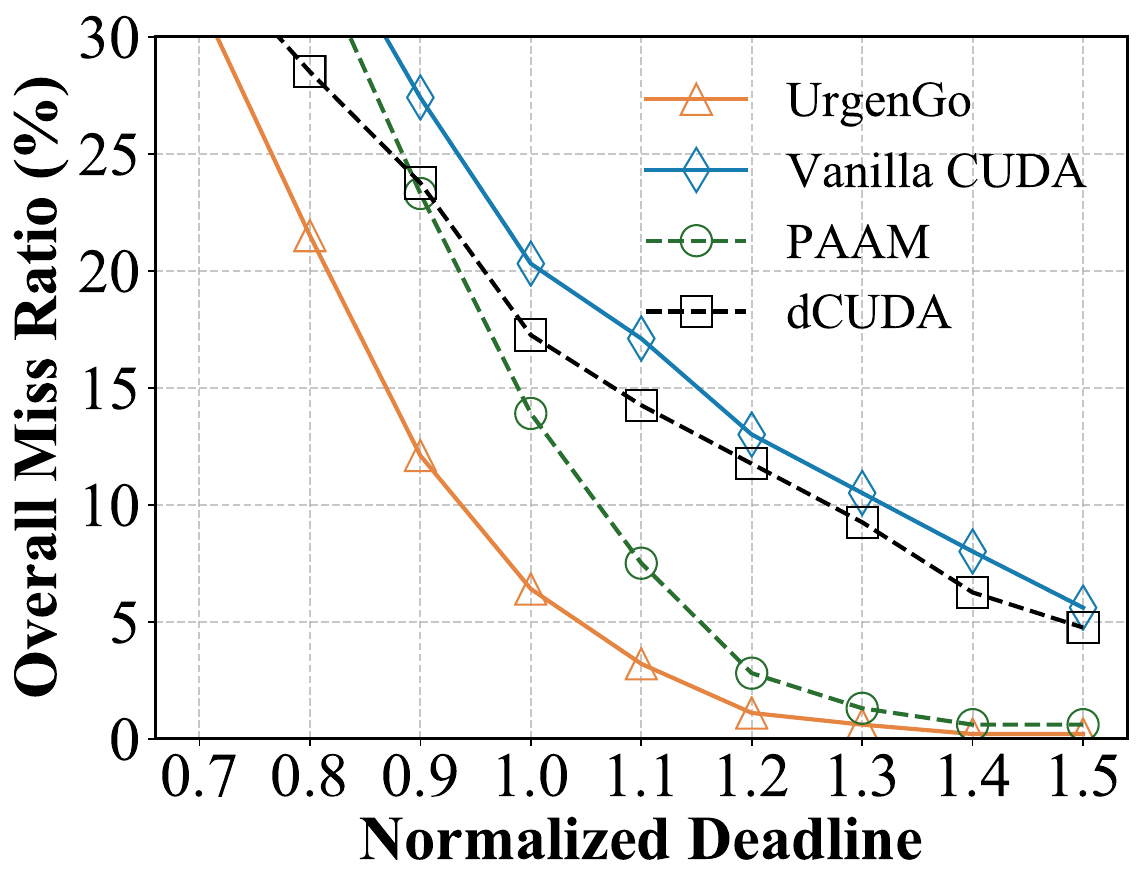}\\
        \captionsetup{width=0.95\linewidth}   \vspace{-12pt}
    \caption{ \small  {Performance with different deadlines.}  } \label{fig:1_overall_ddl} 
    \end{minipage}%
    \begin{minipage}[t]{0.25\linewidth}
        \centering
        \includegraphics[width=0.9\linewidth]{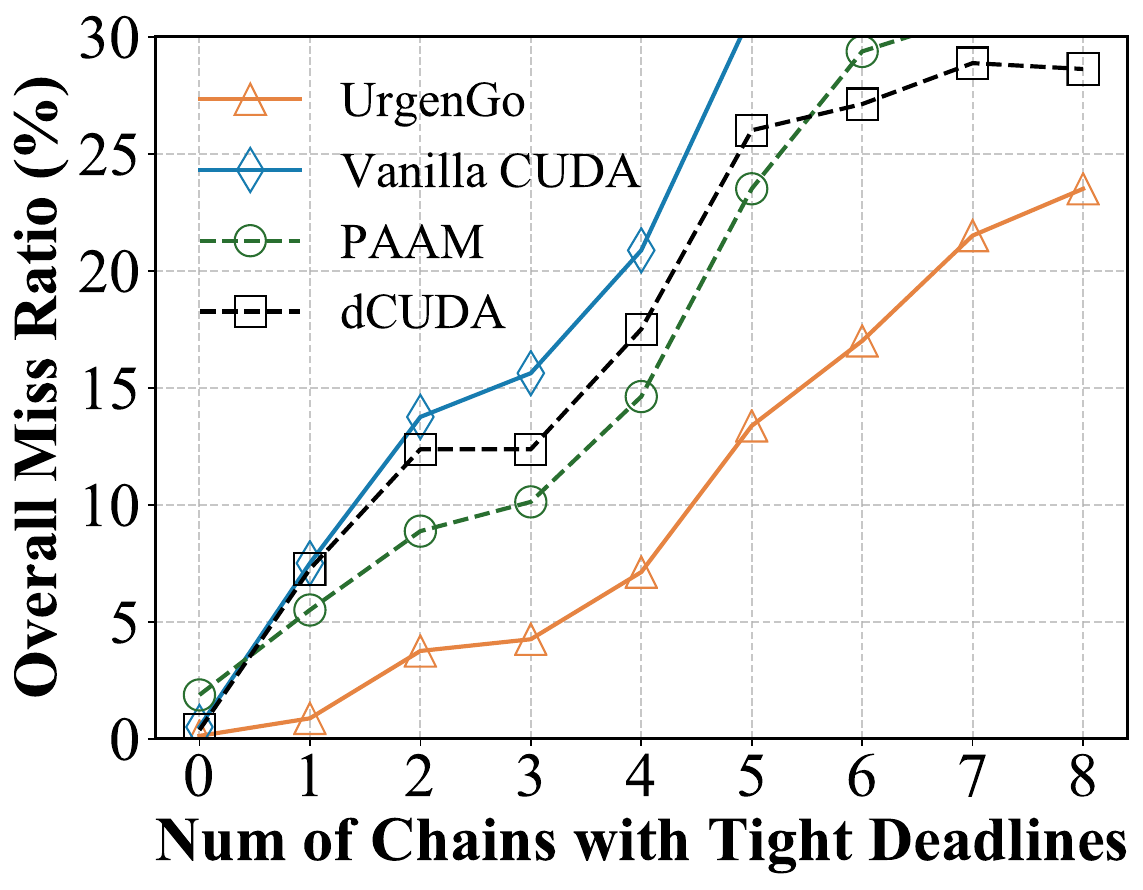}\\
    \captionsetup{width=0.95\linewidth}    \vspace{-12pt}
    \caption{  \small  {Performance with different tightness.} }  \label{fig:2_tightness} 
    \end{minipage}%
    \begin{minipage}[t]{0.25\linewidth}
        \centering
        \includegraphics[width=0.9\linewidth]{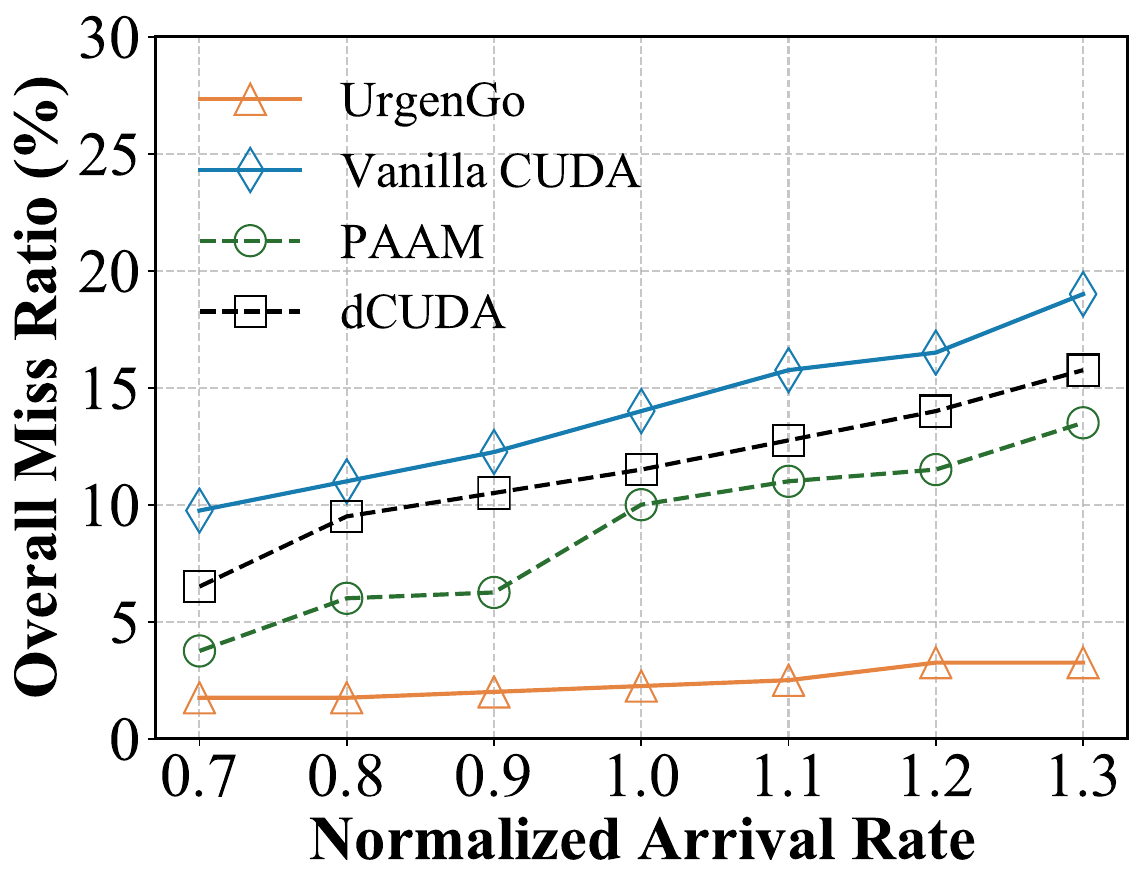}\\
    \captionsetup{width=0.95\linewidth}    \vspace{-12pt}
    \caption{  \small  {Performance on different workflow.} }  \label{fig:3_llama} 
    \end{minipage}%
      \vspace{-6pt}
\end{figure*}
\begin{figure*}[!t]
\centering 
    \begin{minipage}[t]{0.25\linewidth}
       \centering
        \includegraphics[width=0.9\linewidth]{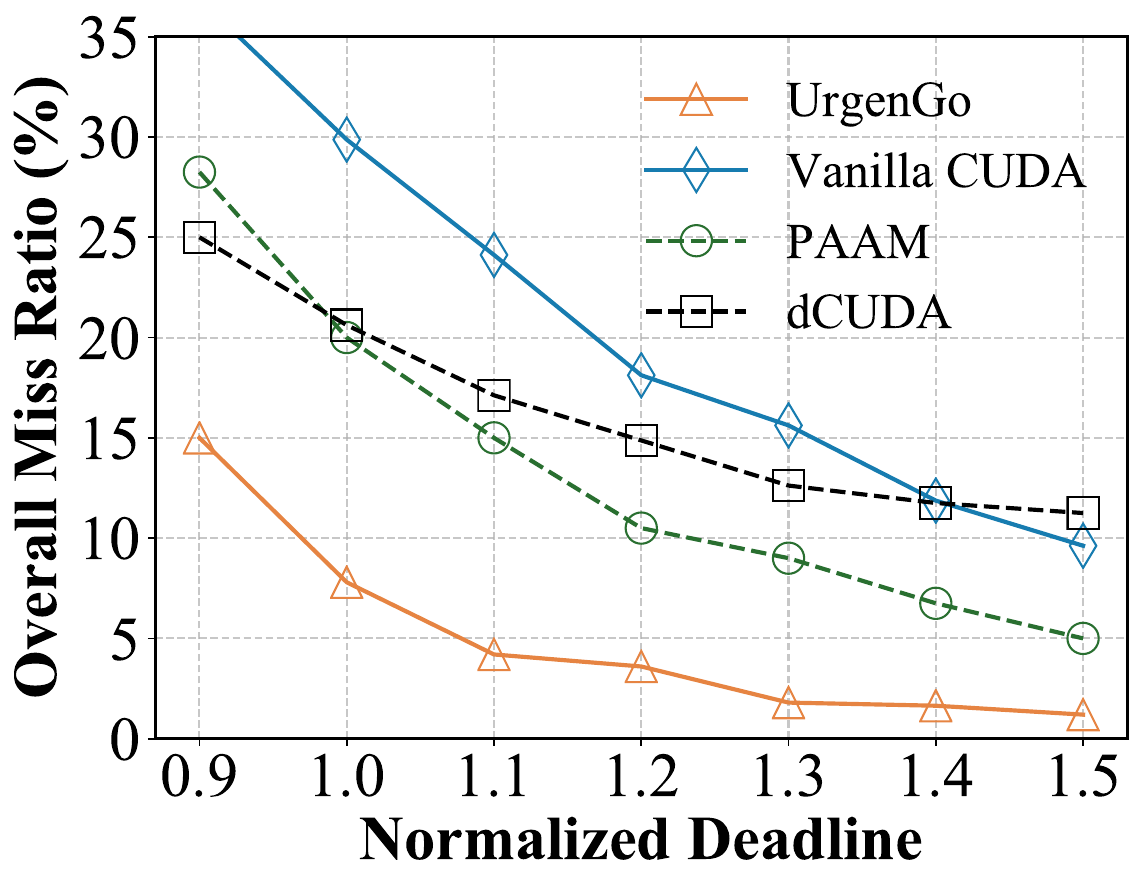}\\
        \captionsetup{width=0.92\linewidth}  \vspace{-12pt}
    \caption{  \small  {Performance on NVIDIA Jetson AGX Orin.}    }  \label{fig:7_overall_ddl} 
    \end{minipage}%
  \begin{minipage}[t]{0.25\linewidth}
        \centering
        \includegraphics[width=0.9\linewidth]{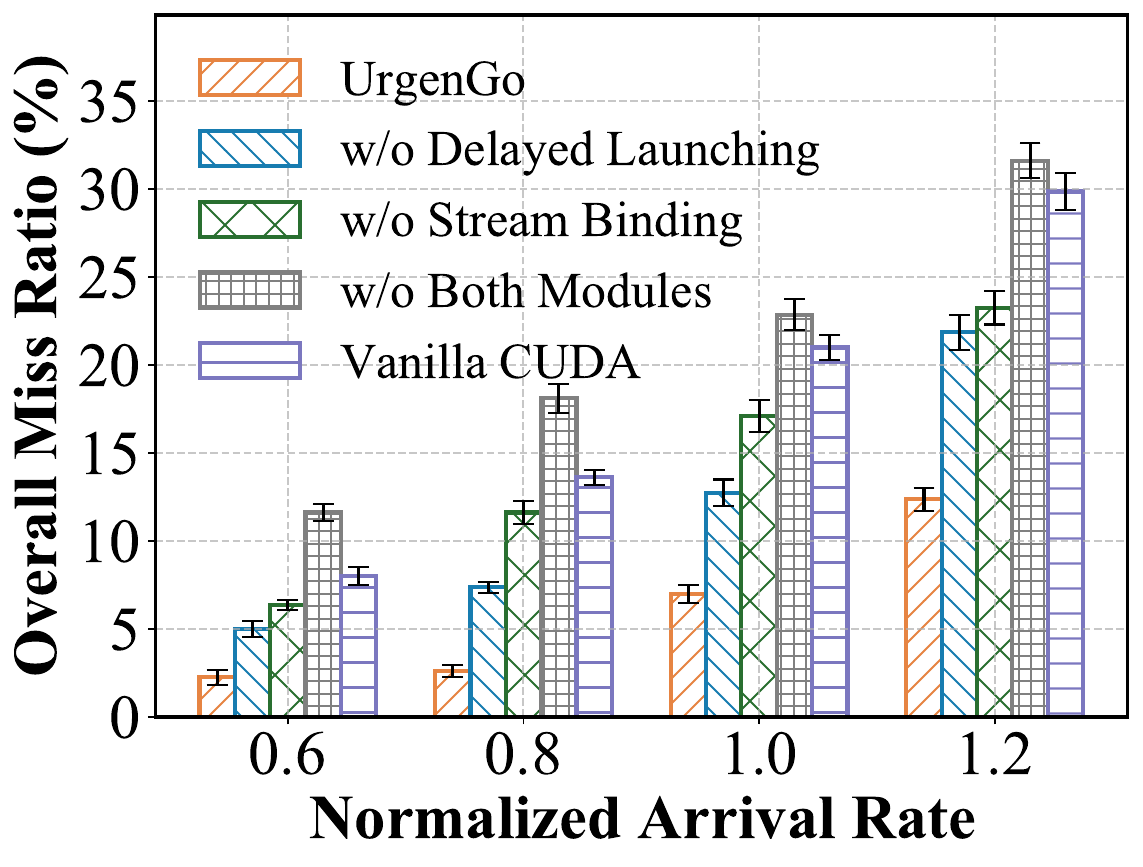}\\
    \captionsetup{width=0.92\linewidth}  \vspace{-12pt}
    \caption{  \small  Ablation studies on key modules.  }  \label{fig:4_ablation}
    \end{minipage}%
    \begin{minipage}[t]{0.25\linewidth}
       \centering
        \includegraphics[width=0.9\linewidth]{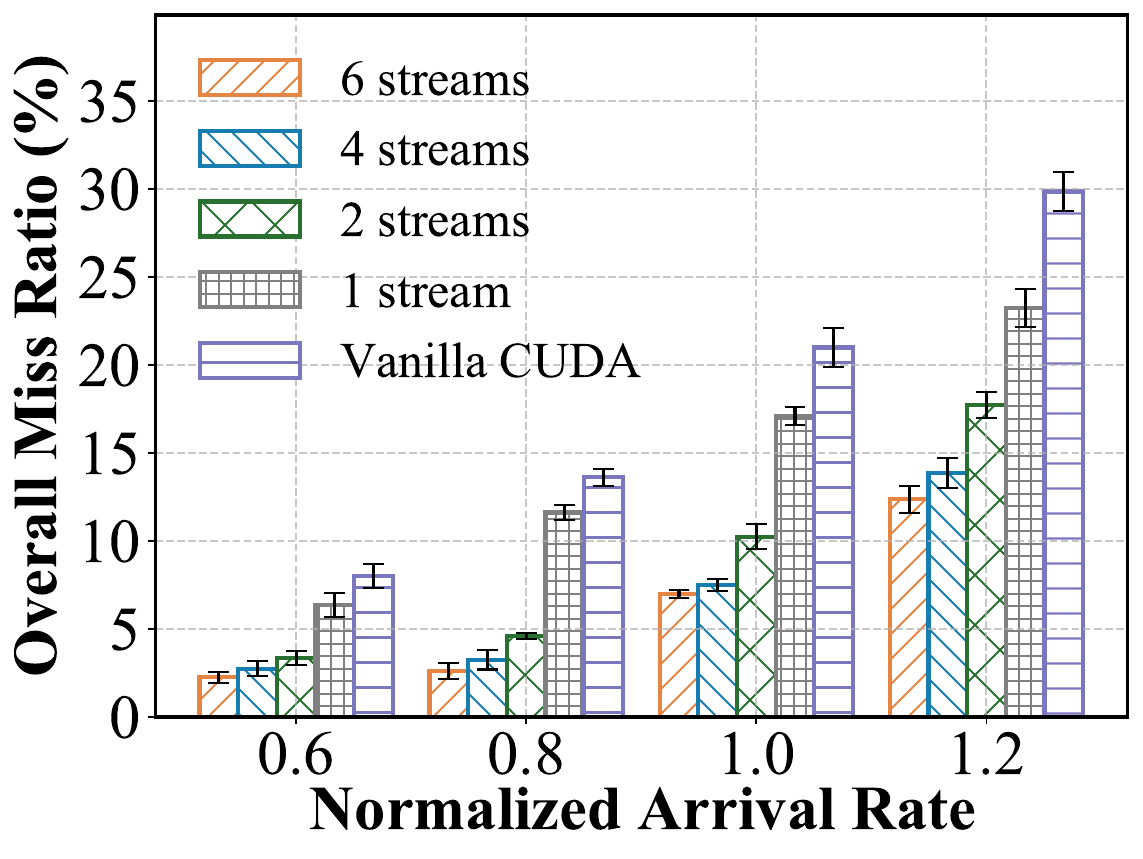}\\
        \captionsetup{width=0.92\linewidth}  \vspace{-12pt}
    \caption{  \small Impact of different stream numbers.    }  \label{fig:5_stream_num} 
    \end{minipage}%
    \begin{minipage}[t]{0.25\linewidth}
       \centering
        \includegraphics[width=0.9\linewidth]{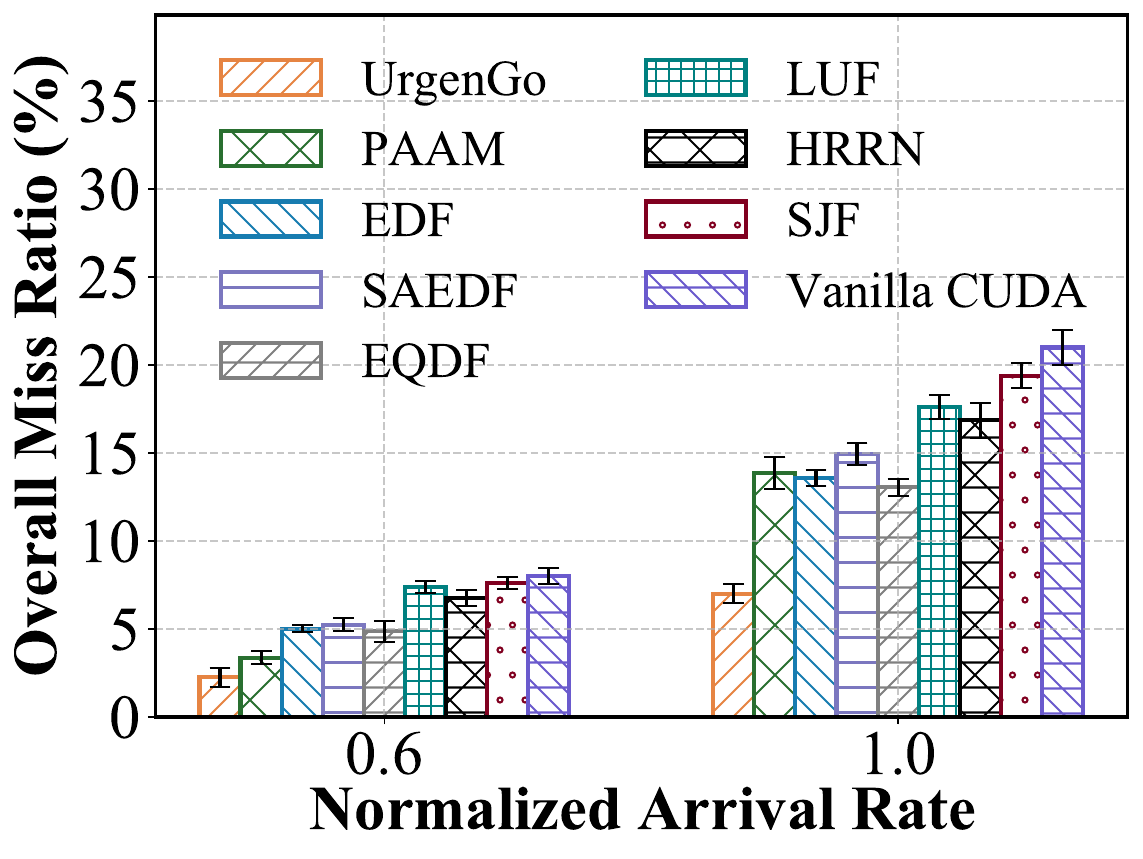}\\
        \captionsetup{width=0.92\linewidth}  \vspace{-12pt}
    \caption{  \small {Impact of different scheduling policies.}    }  \label{fig:6_policy} 
    \end{minipage}%
    \vspace{-16pt}
\end{figure*}

\noindent\textbf{Overall Performance with Different Arrival Rates.}  
In the first experiment, we evaluate how \systemname adapts to varying task arrival rates while keeping $f_{d}$ and $f_{tight}$ fixed.
As shown in Fig.~\ref{fig:0_overall}, performance declines across all methods as the arrival rate increases. However, \systemname consistently outperforms the baselines, demonstrating its robustness even under heavy loads.  
For example, at $f_{a}=0.9$, \systemname achieves a deadline miss rate of 3.8\%, representing relative reductions of 79\%, 61\%, and 76\% compared to vanilla CUDA, PAAM, and dCUDA, respectively. 
We also observe that dCUDA’s performance falls between that of \systemname and vanilla CUDA. This is because dCUDA considers only task utilization and employs a fairness-oriented round-robin scheduling policy, whereas \systemname is both utilization- and deadline-aware, making it better suited for real-time tasks.

\noindent\textbf{Overall Performance with Different Deadlines.}  
In the second experiment, we investigate how \systemname responds to varying deadlines. 
We fix $f_{a}$ and $f_{tight}$ while scaling $f_{d}$. 
As shown in Fig.~\ref{fig:1_overall_ddl}, all methods experience a reduction in deadline miss rates as $f_{d}$ increases, with \systemname continues to outperform the other methods.  
At $f_{d}=1.5$, both \systemname and PAAM maintain miss rates below 1.5\%, significantly outperforming dCUDA and the native GPU scheduler. 
Under a stricter deadline of $f_{d}=1.0$, \systemname achieves a miss ratio of 6.4\%, resulting in a relative reduction of 54\%, 63\%, and 68\%, compared to PAAM, dCUDA, and vanilla CUDA. 
These results emphasize \systemname’s strength in handling kernel collisions, even as deadlines become more flexible.

\noindent\textbf{Overall Performance with Varying Tight Deadline Ratios.}  
{In this experiment, we examine how \systemname responds to different proportions of tight deadline task chains, as some tasks may have stricter deadlines while others allow more flexibility. 
We fix $f_{a}$ and $f_{d}$ while scaling $f_{tight}$. 
As shown in Fig.~\ref{fig:2_tightness}, deadline miss rates increase across all methods as $f_{tight}$ increases. When  $f_{tight}=0$, all methods maintain a deadline miss rate below 1.5\%. 
However, as $f_{tight}$ grows, the performance gap between \systemname and the baseline methods widens. For instance, compared to PAAM, this gap increases from 4.6\% to 12.4\% as $f_{tight}$ 
rises from 10\% to 60\%.}

\noindent\textbf{Overall Performance with Different Workflows.}  
{The previous experiments were conducted using task chains $C_0$ to $C_9$. To further validate the performance of \systemname across different workflows, we evaluate it using task chains $C_6$ to $C_{10}$, which includes an LLM-based task chain. 
Unlike conventional chains, the LLM task chain uses the interval between adjacent token generations as its deadline, based on human reading speed~\cite{liu2024andes} (e.g., 5 tokens/s). We fix $f_{a}$, $f_{d}$ and $f_{tight}$ at their default settings, 
Fig.~\ref{fig:3_llama} shows that \systemname achieves a deadline miss rate below 5\% across different arrival rates and continues to outperform the baseline methods.  
}

\begin{figure*}[!t]
\centering
   \begin{minipage}[t]{0.25\linewidth}
      \centering
        \includegraphics[width=0.9\linewidth]{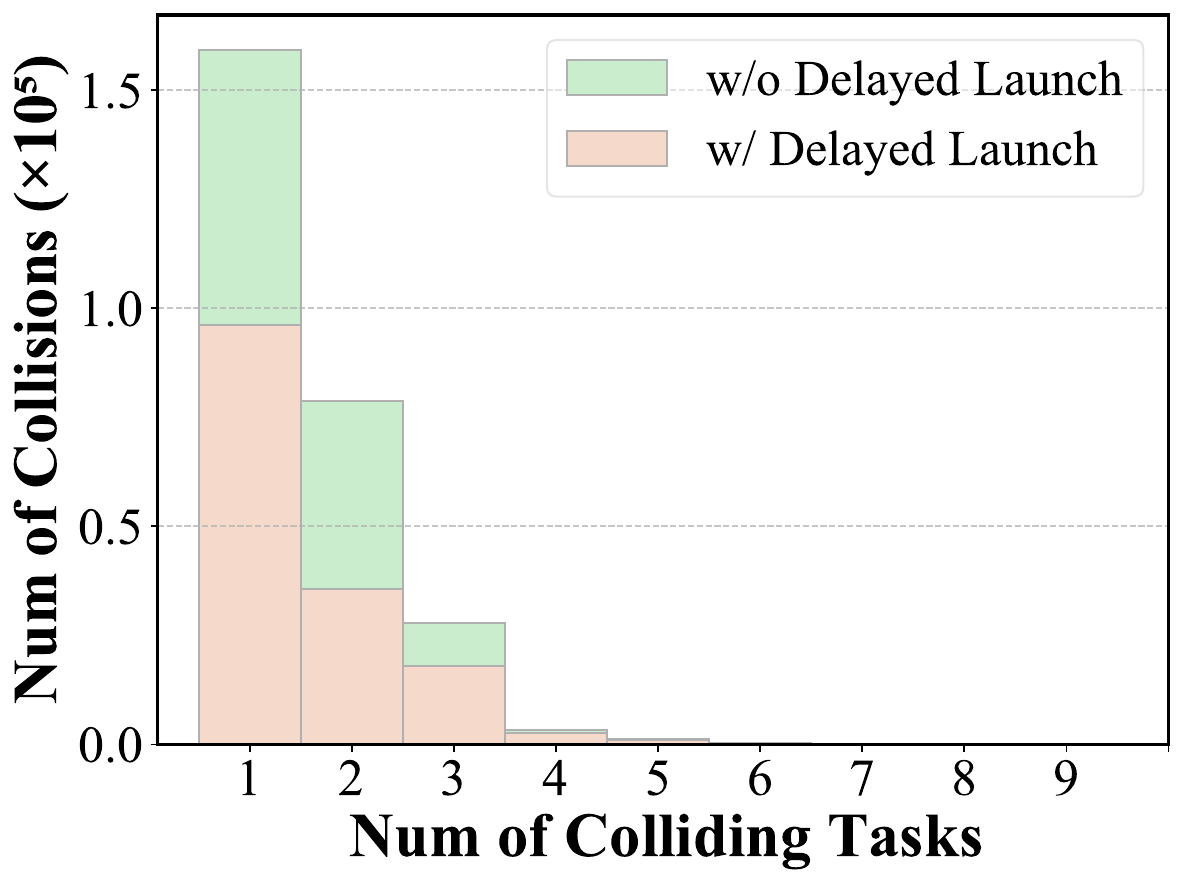}\\
        \captionsetup{width=0.95\linewidth}  \vspace{-12pt}
    \caption{  \small  Impact of delayed launching on collisions.    } \label{fig:8_collision} 
    \end{minipage}%
    \begin{minipage}[t]{0.25\linewidth}
        \centering
        \includegraphics[width=0.9\linewidth]{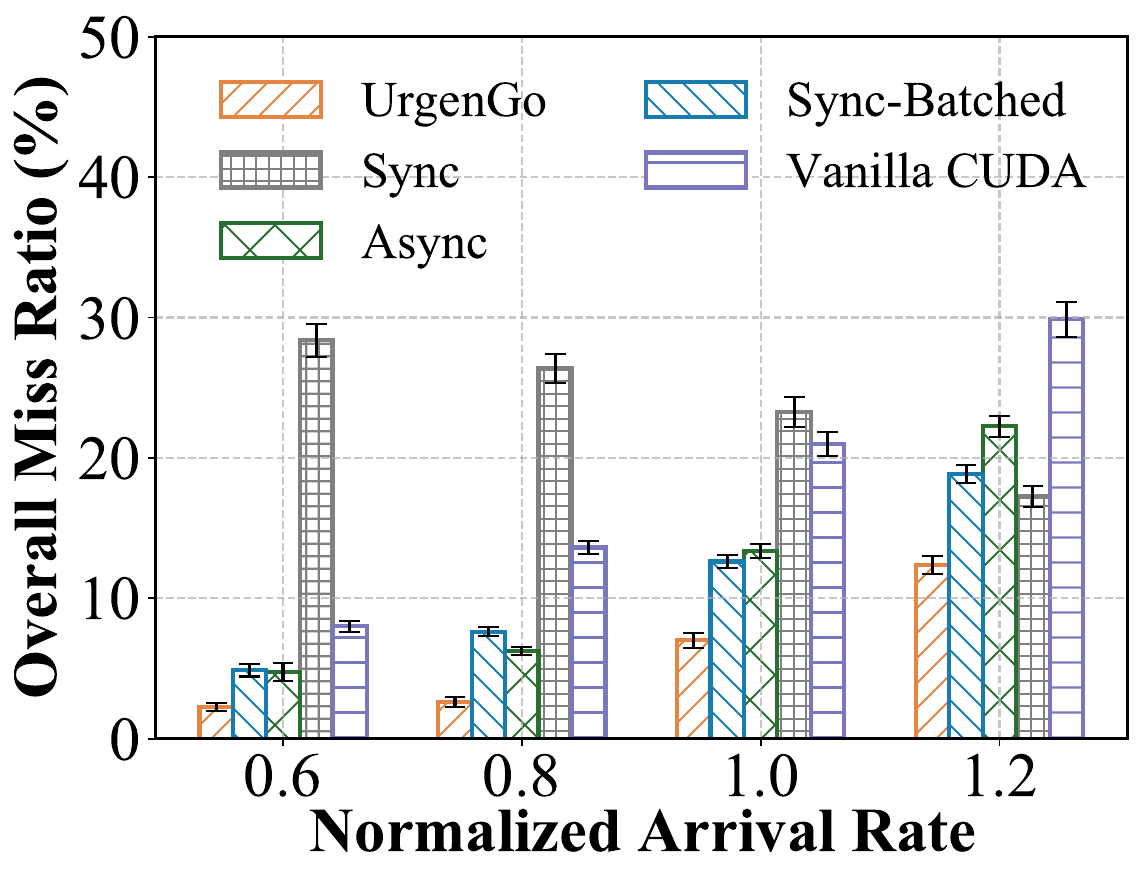}\\
    \captionsetup{width=0.95\linewidth}    \vspace{-12pt}
    \caption{ \small Impact of synchronization mechanisms. } \label{fig:9_launch} 
    \end{minipage}%
    \begin{minipage}[t]{0.25\linewidth}
        \centering
        \includegraphics[width=0.9\linewidth]{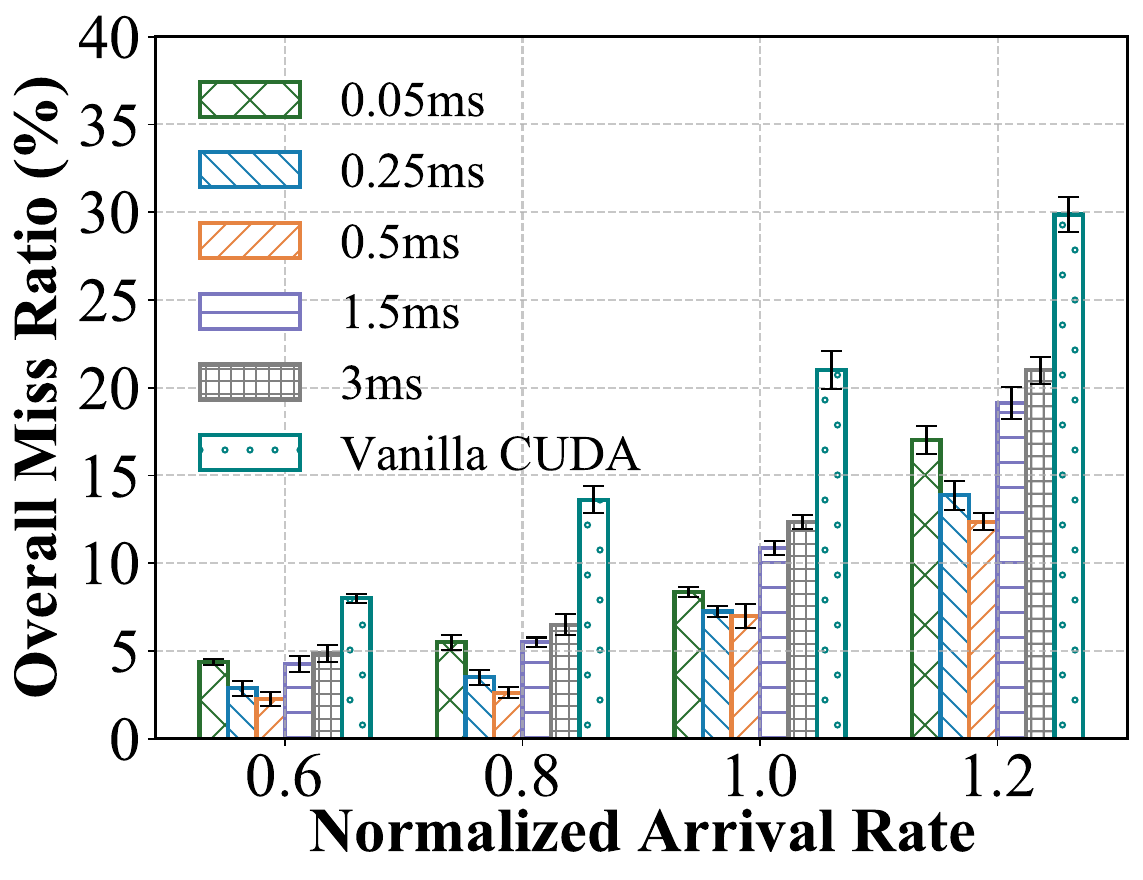}\\
    \captionsetup{width=0.95\linewidth}  \vspace{-12pt}
    \caption{  \small Impact of urgency evaluation intervals.   } \label{fig:10_resolution} 
    \end{minipage}%
    \begin{minipage}[t]{0.25\linewidth}
      \centering
        \includegraphics[width=0.9\linewidth]{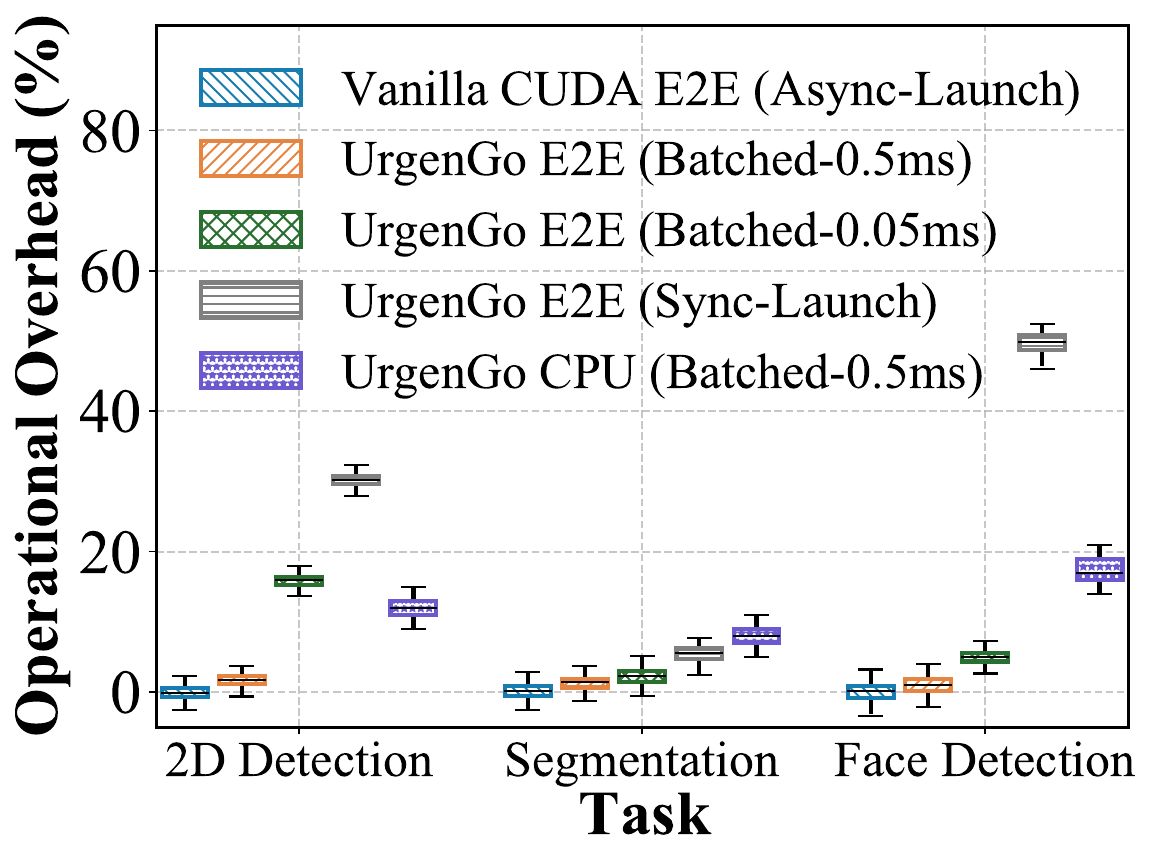}\\
        \captionsetup{width=0.95\linewidth}  \vspace{-12pt}
    \caption{  \small {Operational overhead across different tasks.}    } \label{fig:11_overhead} 
    \end{minipage}%
    \vspace{-8pt}
\end{figure*}

\begin{figure*}[!t]
\centering
    \begin{minipage}[t]{0.25\linewidth}
        \centering
    \includegraphics[width=0.9\linewidth]{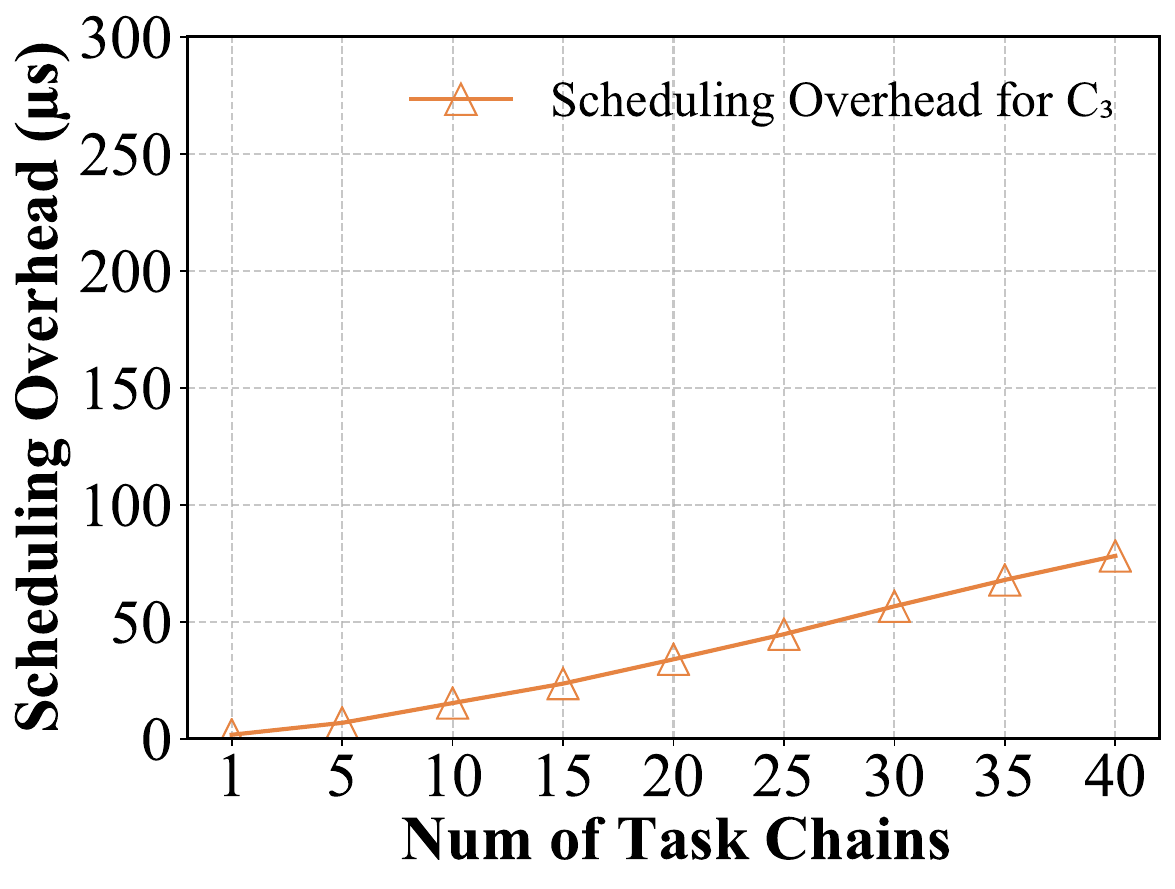}\\
    \captionsetup{width=0.92\linewidth}  \vspace{-12pt}
    \caption{ \small Scheduling overhead for task chain $C_3$.} \label{fig:24_scheduling_overhead}
    \end{minipage}%
    \begin{minipage}[t]{0.25\linewidth}
      \centering
        \includegraphics[width=0.9\linewidth]{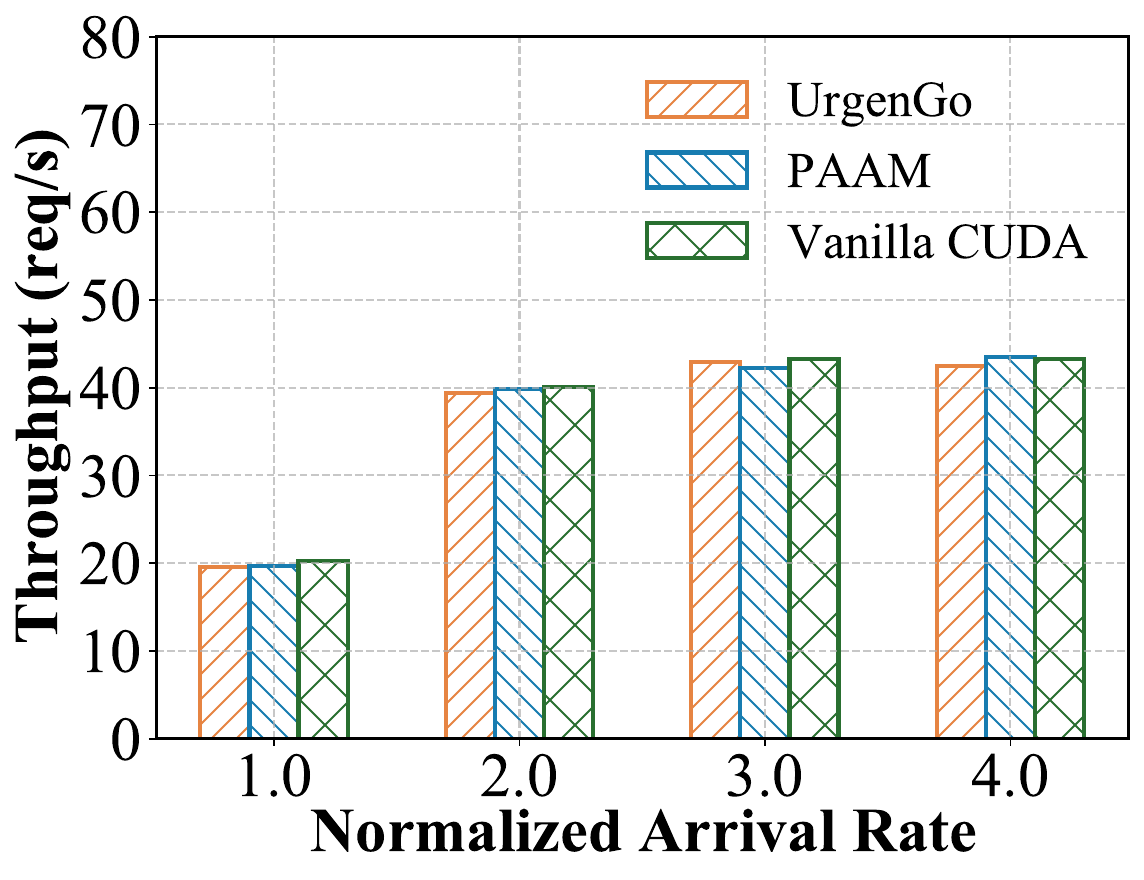}\\
    \captionsetup{width=0.92\linewidth}  \vspace{-12pt}
    \caption{ \small Throughput of different methods.  }  \label{fig:23_throughput}
    \end{minipage}%
    \begin{minipage}[t]{0.5\linewidth}
        \centering
    \includegraphics[width=0.9\linewidth]{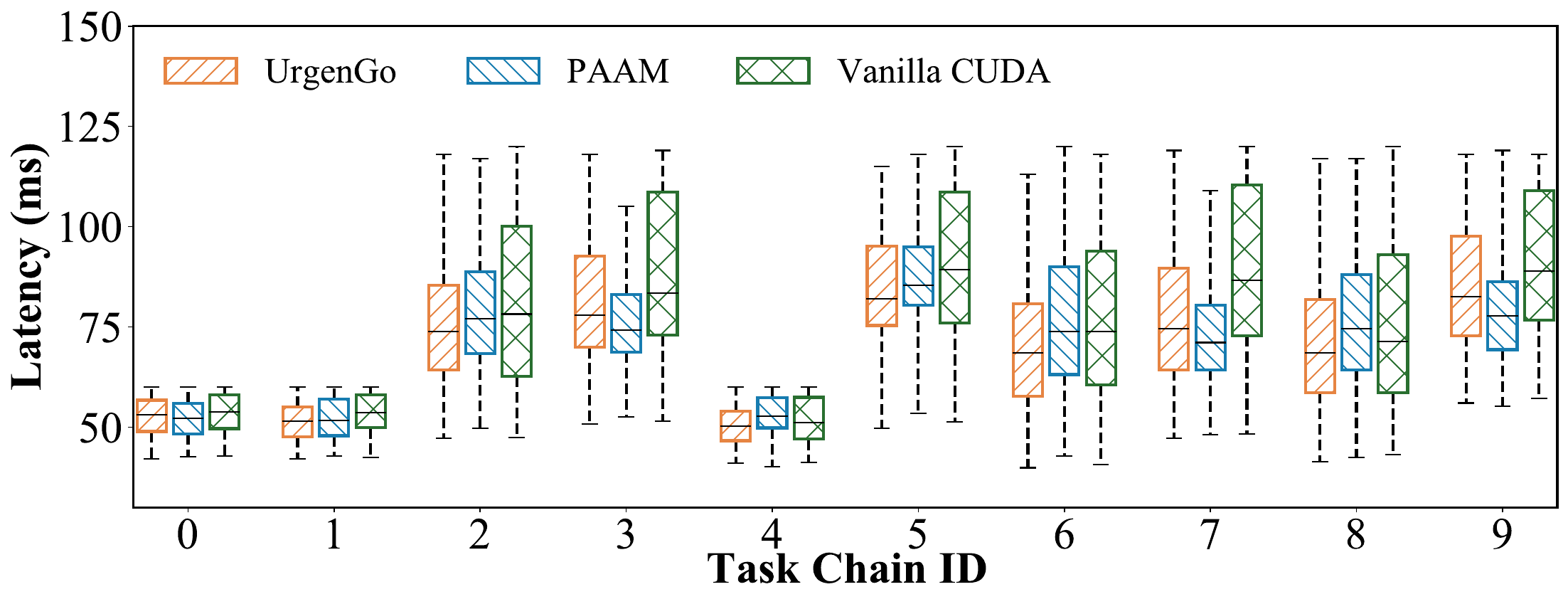}\\
    \captionsetup{width=0.92\linewidth}  \vspace{-12pt}
    \caption{ \small Latency of different task chains.}  \label{fig:25_boxplot}
    \end{minipage}%
     \vspace{-12pt} 
\end{figure*}

\noindent\textbf{Overall Performance with Different Hardware.}  
In this experiment, we evaluate the performance of \systemname on a less powerful embedded device, the NVIDIA Jetson AGX Orin. The Orin platform has a computational capability of 3.3 TFLOPS and 8 SMs, lower than the 21.7 TFLOPS and 46 SMs of the NVIDIA RTX 3070 Ti. To accommodate Orin’s lower computing power, we apply INT8 quantization to all neural networks.  
As shown in Fig.~\ref{fig:7_overall_ddl}, the performance trend is similar to that observed in Fig.~\ref{fig:1_overall_ddl}. 
At $f_{d}=1.0$, \systemname achieves a deadline miss rate of 7.8\%, significantly outperforming the vanilla CUDA's 29.9\%, PAAM's 20.1\%, and dCUDA's 20.6\%. 

\vspace{-8pt}
\subsection{Exploring \systemname Design Options} 
Next, we examine the core modules of \systemname to quantify their contributions and evaluate key design options. 
This section uses the default settings of $f_{tight}=40\%$ and $f_{d}=1.0$ to ensure consistent comparison across experiments. 
 
\vspace{2pt}\noindent\textbf{Stream Binding and Delayed Launching.} 
To understand the independent and combined impact of stream binding and delayed launching, we evaluate them in isolation and together.
Fig.~\ref{fig:4_ablation} shows that when $f_{a}=1.0$, delayed launching alone reduces the deadline miss rate by 5.7\%, while stream binding reduces it by 10.1\%. When used together, they achieve a 15.8\% reduction. This confirms that these techniques complement each other—delayed launching mitigates kernel collisions, while dynamic stream binding efficiently allocates stream priorities based on task urgency.

\vspace{0pt} \noindent\textbf{Number of Binding Streams.}
We also examine the effect of varying the number of available streams on performance, as shown in Fig.~\ref{fig:5_stream_num}. As expected, increasing the number of streams from 1 to 6 consistently reduces the deadline miss rate. The most significant drop, 7.1\%, occurs when increasing from 1 to 2 streams, because this avoids the most frequent cases when two tasks collide. This highlights the importance of having multiple binding streams to reduce task contention, though the benefit diminishes after reaching a threshold where additional streams provide little further improvement.

\vspace{0pt}\noindent\textbf{Comparison of different Policies.}
{ To assess the necessity of \systemname's urgency-based policy, we compare it against alternative approaches, including EDF-family such as EDF, SAEDF~\cite{SAEDF}, and EQDF~\cite{SAEDF}, utilization-based policy such as Lowest Chain Utilization First~\cite{anderson2005edf}, as well as SJF-family such as SJF and Highest Response Ratio Next (HRRN).
Fig.~\ref{fig:6_policy} shows that \systemname consistently outperforms all other policies. When $f_{a}$ is 1.0, EQDF achieves the best performance among the baseline policies with a deadline miss rate of 13.05\%, while \systemname achieves 7\%. This result highlights that urgency-based policy leads to better prioritization for multi-task chains under hybrid CPU and GPU execution. 
}

\vspace{0pt}\noindent\textbf{Delayed Launching Schemes.} 
To investigate how delayed launching reduces collisions, we track the number of kernel collisions for urgent kernels in Fig.~\ref{fig:8_collision}. Most collisions occur with two or three active tasks. Delayed launching reduces collisions by 41.1\%, 55.7\%, 46.0\%, and 22.1\% for 2, 3, 4, and 5 colliding tasks, respectively. This indicates that delayed launching is particularly effective in managing moderate contention scenarios, typical in multi-task environments.

\vspace{0pt}\noindent\textbf{Kernel Launch Synchronization Mechanisms.}
We compare kernel launch synchronization mechanisms to understand how \systemname minimizes synchronization delays.
Fig.~\ref{fig:9_launch} shows that \systemname consistently outperforms synchronous and asynchronous approaches. At $f_{a}=1.0$, \systemname achieves a 7.0\% miss rate, reducing miss rates by 5.6\%, 6.3\%, and 16.2\% compared to synchronous batched launching, asynchronous launching, and synchronous launching, respectively. 
This demonstrates that \systemname's delayed launching and dynamic urgency evaluation effectively balance synchronization overhead and task prioritization.

\vspace{0pt}\noindent\textbf{Urgency Evaluation Interval $\Delta_{eval}$.}
We then investigate the impact of different urgency evaluation intervals, $\Delta_{eval}$, on system performance in Fig.~\ref{fig:10_resolution}. 
Small intervals introduce excessive overhead, while large intervals delay task prioritization updates. At $f_{a}=1.0$, a 0.5~ms interval achieves the optimal balance, reducing the miss rate by up to 5.3\% compared to other intervals. This shows that \systemname's default urgency evaluation settings provide a well-balanced trade-off between evaluation accuracy and synchronization overhead.

\subsection{Evaluation of System Overhead} 
\vspace{0pt}\noindent\textbf{Operational Overhead.}       
We first present the operational overhead of \systemname in Fig.~\ref{fig:11_overhead} and Tab.~\ref{tab:basic}, with each task executed in isolation. Fig.~\ref{fig:11_overhead} reports results for neural network–based tasks, while Tab.~\ref {tab:basic} shows results for basic CUDA operations.
Compared to vanilla CUDA (1.0$\times$), \systemname -- using batched launching and a 0.5~ms urgency evaluation interval -- increases execution time by only 1\% across all three tasks, indicating minimal overhead.
The segmentation task is less affected by the launching method due to its lower number of kernel invocations (Tab.~\ref{profiling}).
We also report CPU-side overhead, which is about 17\% compared to vanilla CUDA. Given that CPU launching time is typically small -- e.g., 7~ms for 2D detection and 2~ms for face detection --  and can often be hidden through asynchronous execution, this overhead remains acceptable. 
Basic CUDA operations follow a similar trend. 
Notably, we include the cudaGetDevice API in Tab.~\ref{tab:basic} to explicitly measure API interception cost. cudaGetDevice is a lightweight query requiring neither computation nor memory copying, and thus isolates the interception cost from kernel manipulation.
Results show that API interception overhead is negligible at the microsecond level. 
Overall, the small operational overhead validates the efficiency of our custom GPU libraries.

\begin{figure*}[!t]
\centering
    \begin{minipage}[t]{0.25\linewidth}
        \centering
    \includegraphics[width=0.9\linewidth]{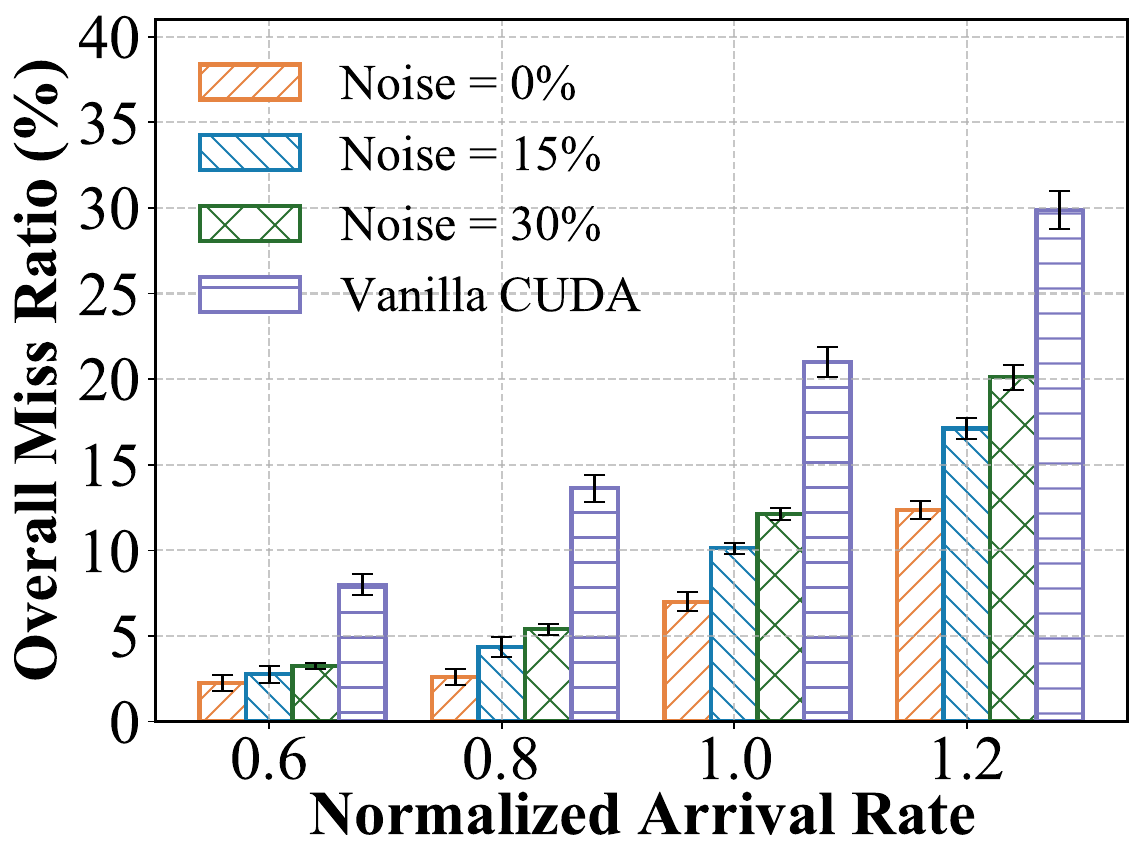}\\
    \captionsetup{width=0.92\linewidth}  \vspace{-12pt}
    \caption{ \small Impact of urgency evaluation error.}  \label{fig:12_noisy}
    \end{minipage}%
    \begin{minipage}[t]{0.25\linewidth}
      \centering
        \includegraphics[width=0.9\linewidth]{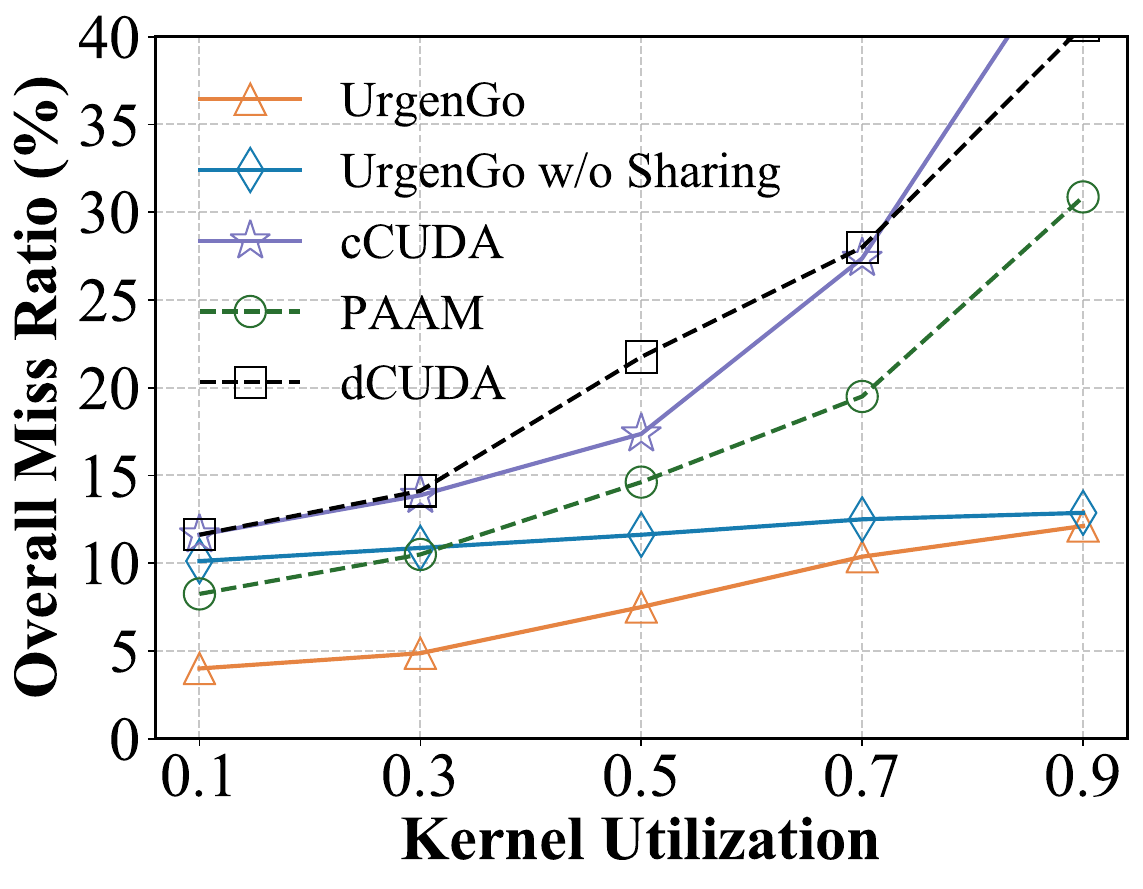}\\
    \captionsetup{width=0.92\linewidth}  \vspace{-12pt}
    \caption{ \small {Impact of GPU utilization of kernels.}  }  \label{fig:13_vector_util}
    \end{minipage}%
    \begin{minipage}[t]{0.25\linewidth}
    \centering
        \includegraphics[width=0.9\linewidth]{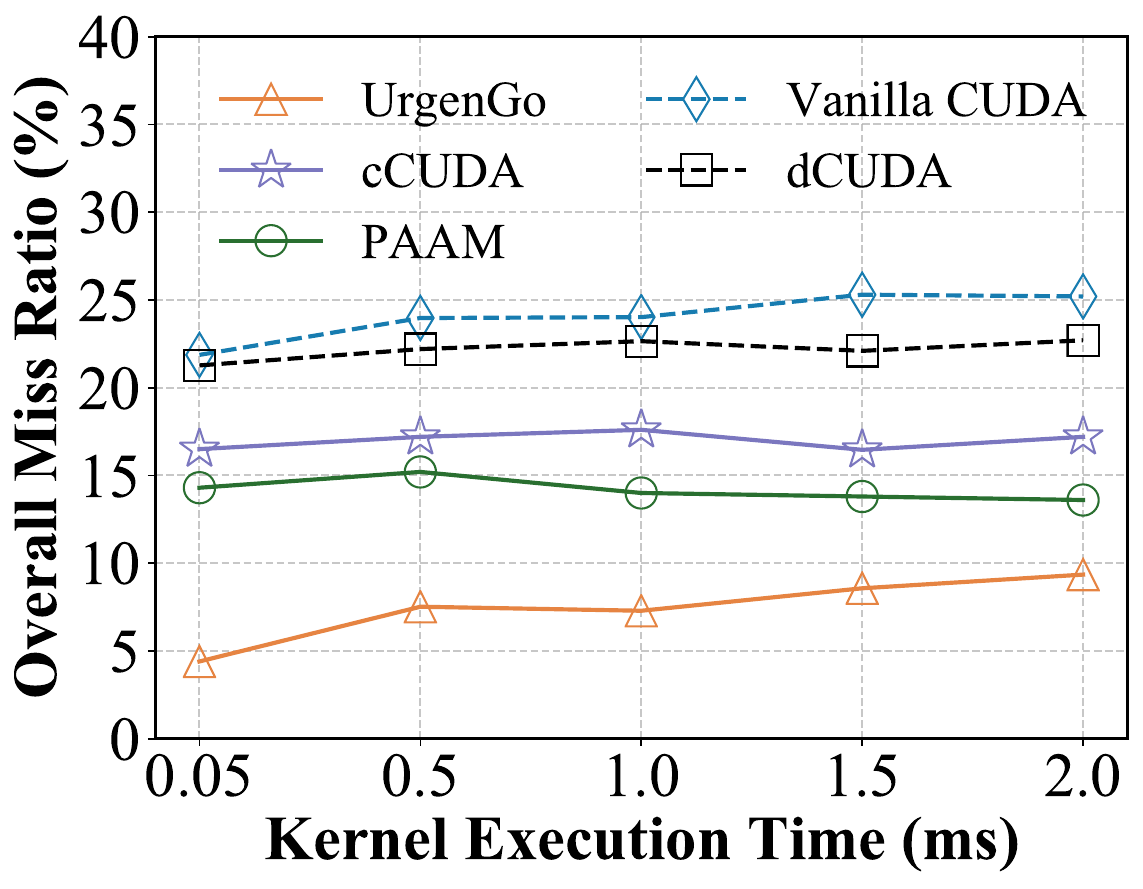}\\
        \captionsetup{width=0.92\linewidth}  \vspace{-12pt}
    \caption{ \small {Impact of the execution time of kernels.}}  \label{fig:14_vector_time} 
    \end{minipage}%
    \begin{minipage}[t]{0.25\linewidth}
          \centering
        \includegraphics[width=0.9\linewidth]{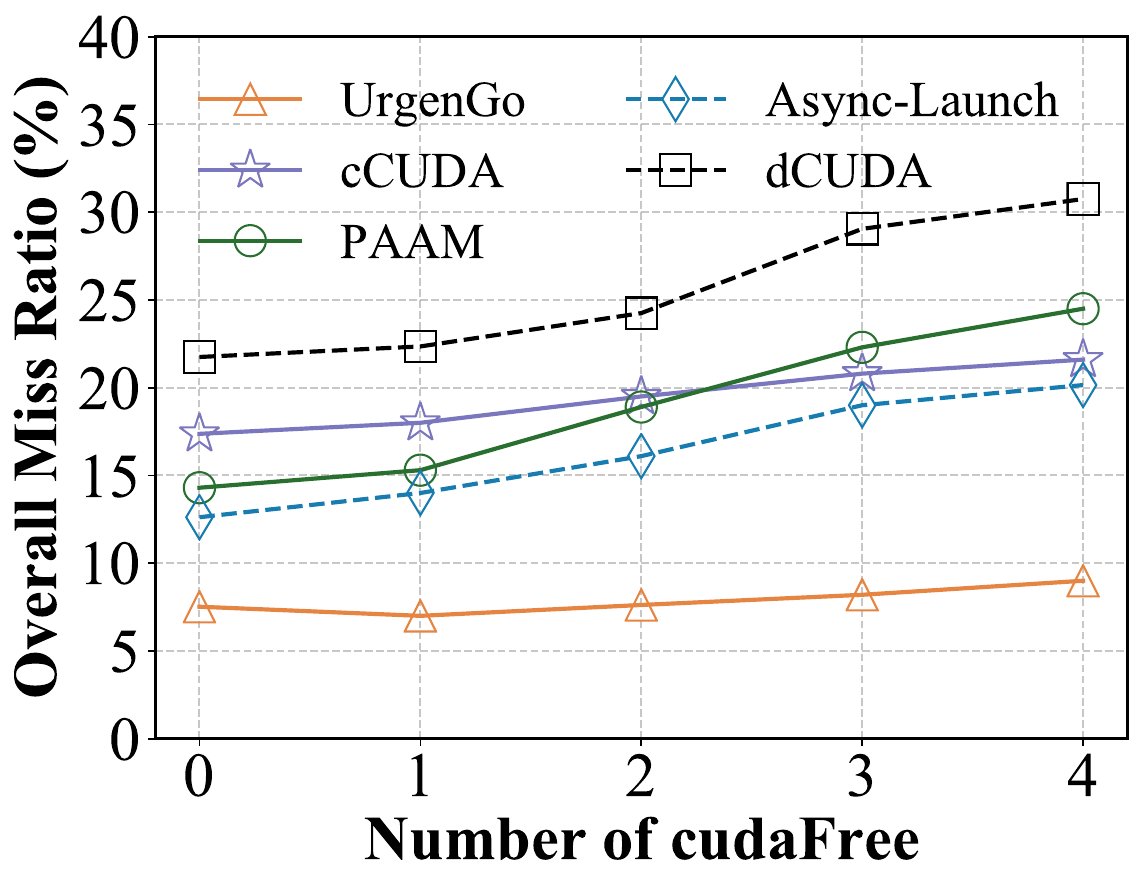}\\
        \captionsetup{width=0.92\linewidth}  \vspace{-12pt}
    \caption{ \small {Impact of the number of cudaFree calls.}}  \label{fig:15_vector_free} 
    \end{minipage}%
    \vspace{-12pt}
\end{figure*}

\vspace{0pt}\noindent\textbf{Scheduling Overhead.}  
To further evaluate the implementation of our urgency-based scheduler, we measure the accumulated scheduling overhead for task chain $C_3$ throughout its execution. Unlike operational overhead, which depends mainly on the number of kernels within a task, scheduling overhead is determined primarily by the total number of task chains due to the scheduler’s $O(N)$ complexity.  
As shown in Fig.~\ref{fig:24_scheduling_overhead}, the scheduling overhead scales linearly with the number of task chains. Specifically, when there are 20 task chains, the total scheduling overhead is $34\,\mu\text{s}$ -- only 0.07\% of $C_3$’s 48~ms execution time  -- indicating a negligible impact.
Overall, these results confirm that the scheduling overhead of \systemname remains minimal across various scenarios.
 
\vspace{0pt}\noindent\textbf{Overall Throughput.}     
We then present the overall throughput. The experiment is conducted on four task chains, each configured identically to task chain $C_3$. All task chains are executed without deadlines to better isolate the effect of \systemname on throughput.
As shown in Fig.~\ref{fig:23_throughput}, at an arrival rate of 1.0, \systemname achieves a throughput of 19.60 req/s, slightly below PAAM’s 19.66 and vanilla CUDA’s 20.12. At an arrival rate of 4.0, \systemname reaches 42.45 req/s, compared to 43.35 for PAAM and 43.22 for vanilla CUDA.
Although \systemname incurs a slight throughput reduction due to its operational overhead, the degradation remains modest -- under 2.6\%.

\vspace{0pt}\noindent\textbf{Latency for Each Task Chain.}  
We include additional latency metrics to further illustrate \systemname's performance. As shown in Fig.\ref{fig:25_boxplot}, with $f_{tight}=30\%$, $f_{d}=1.0$, and $f_{a}=1.0$, we report the latency of each task chain for \systemname, PAAM, and vanilla CUDA. \systemname achieves the lowest average latency at 74.0~ms, compared to 74.7~ms for PAAM and 78.7~ms for vanilla CUDA. While \systemname is not explicitly designed to minimize task-chain latency, its dynamic prioritization strategy reduces kernel execution collisions, thereby indirectly lowering the latency.
 
 \begin{table}[b] \vspace{-8pt}
         \setlength\tabcolsep{2pt} 
        \renewcommand\arraystretch{1}
           \centering \scriptsize
        	\begin{tabular}{|c|c|c|c|c|}
                 \hline  
                 \multirow{2}{*}[-3.5ex]{\makecell{\thead{Operation}}}  & \multicolumn{2}{c|}{   \systemname   } & \multicolumn{2}{c|}{  Vanilla CUDA    }   \\  
    			    \cline{2-5} 
    		        &  \thead{ \scriptsize{ End-to-End  } \vspace{-3pt} \\ \scriptsize{Latency (\textmu s )}}     &   \thead{ \scriptsize{ Execution Time    } \vspace{-3pt} \\ \scriptsize{ on CPU (\textmu s )}}   &  \thead{ \scriptsize{ End-to-End  } \vspace{-3pt} \\ \scriptsize{Latency (\textmu s )}}   &   \thead{ \scriptsize{  Execution Time  } \vspace{-3pt} \\ \scriptsize{  on CPU (\textmu s )}}    \\		
    		          \hline  
                         vectorAdd     &   3341 ($\pm$ 98)       & 1152 ($\pm$ 43)      &  3312 ($\pm$ 124)     &  993 ($\pm$ 35)      \\  
                         \hline 
                            matrixMult      &  377 ($\pm$ 47 )      &  229 ($\pm$ 45  )      &  374  ($\pm$ 47 )     &  224 ($\pm$ 37 )    \\  
        			    \cline{2-5}
                         \hline 
                            cudaFree      &   191 ($\pm$ 42)     &   188 ($\pm$ 41)       &  183 ($\pm$ 41)       &  180 ($\pm$ 41)       \\  
        			  \hline 
                            cudaGetDevice      &   {4.16 ($\pm$ 1.24)}     &   1.19 ($\pm$ 0.16)      &  3.77 ($\pm$ 1.08)      &  1.16 ($\pm$ 0.07)       \\  
        			  \hline  
        			\end{tabular} 
         \caption{ {Operational overhead for basic operations.}} \vspace{-2pt}
           \label{tab:basic}
        \end{table}

\vspace{-8pt} 
\subsection{Evaluation of Workload Parameters}
In this section, we study the impact of key workload parameters on the performance of \systemname, helping us understand its robustness under varied real-world conditions.  

\vspace{2pt}\noindent\textbf{Impact of Task Urgency Estimation Error.} 
Understanding how sensitive \systemname is to errors in task urgency estimation is important, as real-world systems face uncertainty in task execution. To evaluate this, we introduce noise to the estimated execution times used to calculate task urgency, with noise uniformly sampled for each task instance (Fig.~\ref{fig:12_noisy}). 
Even with 30\% of added noise, \systemname maintains an 8.9\% advantage over PAAM at $f_{a}=1.0$. This demonstrates that \systemname is resilient to imperfect urgency estimations, which is critical to support real-world deployment where exact task execution times are difficult to predict.

\noindent\textbf{Impact of Kernel GPU utilization.} 
Since high GPU utilization is common in resource-constrained environments, it’s important to see how \systemname handles increasing utilization without compromising scheduling performance.
In Fig.~\ref{fig:13_vector_util}, we replace half of the GPU tasks with custom operations (vector addition~\cite{rodinia} and histogram computation~\cite{rodinia}) to control execution time and utilization, enabling fair comparisons with cCUDA. We then incrementally increase the utilization levels of these custom kernels. 
As utilization rises, the deadline miss rate of \systemname increases from 4.1\% to 12.1\%. However, \systemname consistently outperforms all other methods across different utilization levels, indicating its effectiveness in managing heavy GPU workloads.   
By contrast, cCUDA focuses solely on maximizing overall system utilization without considering per-task time constraints, limiting its suitability for real-time workloads.

\noindent\textbf{Impact of Kernel Execution Time.} 
Since kernel execution times vary significantly across workloads, it is crucial to study their impact on task performance.
In Fig.~\ref{fig:14_vector_time}, we increase the execution time of custom kernels while maintaining a constant total task execution time by reducing the number of kernels.
As kernel execution times grow from 0.05~ms to 2.0~ms, \systemname’s miss rate increases by 4.9\%, primarily due to longer kernel collisions. This result highlights a trade-off in task granularity -- longer kernel durations lead to higher contention and reduced overall efficiency.

\noindent\textbf{Performance with cudaFree.}
cudaFree calls introduce global synchronization~\cite{pitfall}, which can significantly impact system performance, especially under heavy workloads.  
To evaluate \systemname's ability to handle this overhead,
we change the number of cudaFree calls across tasks in Fig.~\ref{fig:15_vector_free}. 
When 4 tasks invoke cudaFree, PAAM’s miss rate drops from 14.3\% to 24.5\%, and asynchronous launching drops from 12.6\% to 20.1\%. In contrast, \systemname sees only a minor decline from 7.5\% to 9.0\%. demonstrating its resilience to global synchronization, whether introduced by frequent memory deallocation or other synchronization operations.

\vspace{-8pt}
\subsection{Long Term Evaluation} 
In this section, we study the long-term behavior of \systemname to understand how it performs over extended periods of operations.  
We conducted a two-hour live experiment using the default settings from the previous section.
 
\begin{figure}[!t]
\centering  
    \begin{minipage}[t]{\linewidth} \centering 
         \includegraphics[width=\linewidth]{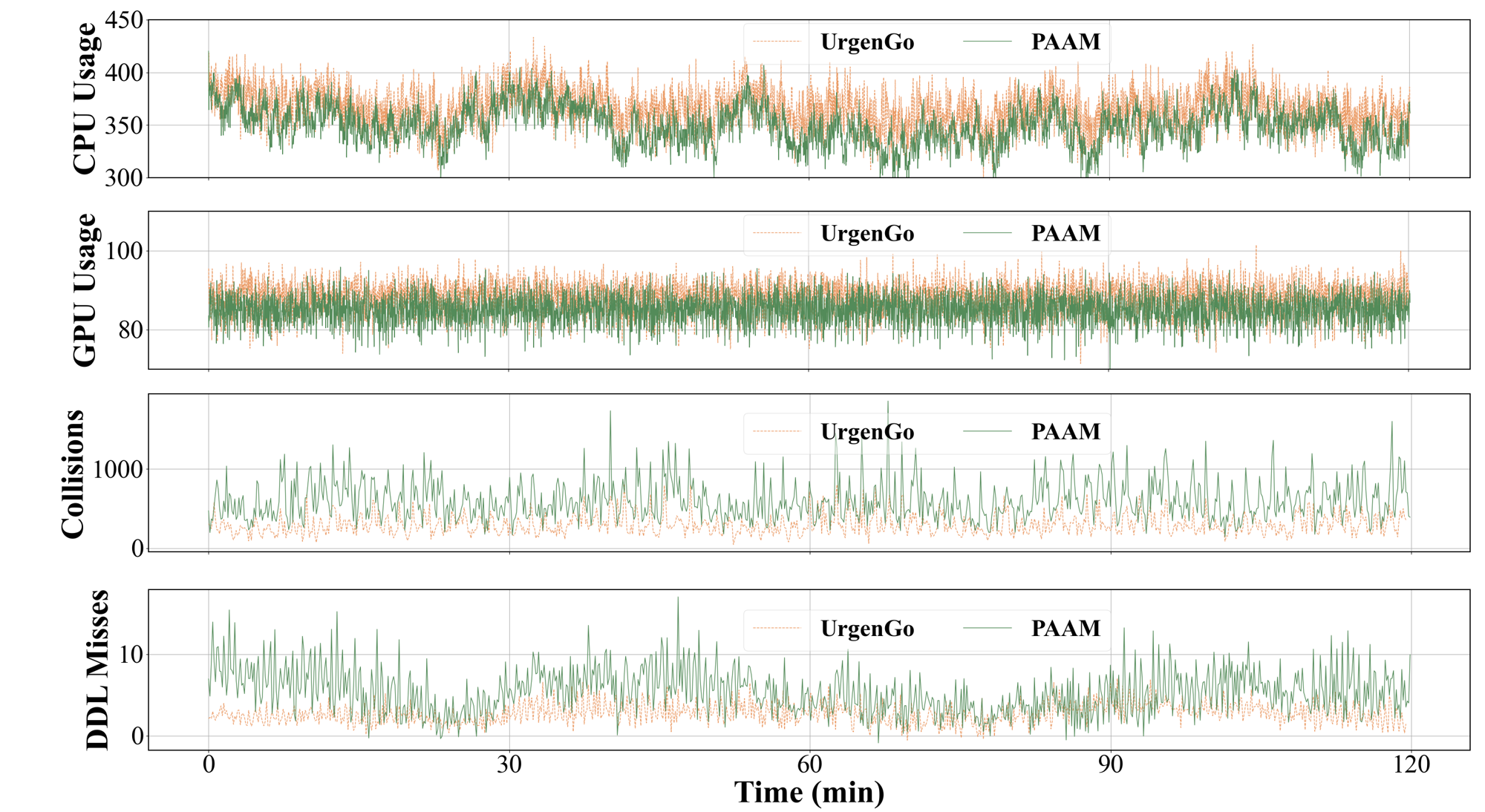}\\
        \captionsetup{width=0.98\linewidth} 
         \vspace{-6pt}
         \caption{ \small Runtime statistics collected over a two-hour period, with the X-axis representing time in minutes.   } \label{fig:threeoverhead} 
    \end{minipage}%
    \vspace{-8pt}
\end{figure}

\noindent\textbf{Runtime Resource Utilization.}  
Fig.~\ref{fig:threeoverhead}(a)(b) tracks CPU and GPU utilization averaged over one-second intervals. 
\systemname shows 4.5\% higher CPU usage and 4.8\% higher GPU usage than PAAM. This increase reflects the higher number of tasks successfully scheduled on both the CPU and GPU, preventing idle cycles.
In contrast, PAAM's higher rate of missed deadlines results in early task termination, leading to reduced resource utilization. This observation shows that \systemname, with its sustained higher utilization, can better handle continuous, real-time workloads without bottlenecks.

\noindent\textbf{Runtime Kernel Collision Avoidance.}  
Fig.~\ref{fig:threeoverhead}(c)(d) presents the number of kernel collisions and task deadline misses over time. 
\systemname consistently experiences fewer kernel collisions and a lower deadline miss rate than PAAM, with average reductions of 51\% and 48\%, respectively, highlighting its effectiveness of task prioritization. Additionally, we observe that both collisions and deadline misses show periodic peaks and valleys, corresponding to synchronized sensor input cycles. 
Interestingly, these collisions and deadline misses are not strictly correlated, suggesting an opportunity to further optimize \systemname by focusing on mitigating only those collisions that lead to deadline misses.

\vspace{-6pt}
\section{Related Work  } 
\noindent\textbf{Task Scheduling for Mobile/Edge Applications. } 
Many techniques have been proposed for scheduling mobile applications:  
RT-mDL~\cite{ling2021rt} jointly considers DNN model scaling and scheduling at the task level.  
Miriam and nnJIT~\cite{mahajan2023better,zhao2023miriam} leverage compile-time optimizations at the kernel level, improving their performance on edge devices.  
Soar~\cite{shi2024soar} mitigates resource contention from multiple concurrent tasks through edge-node collaboration. 
Similarly, AccuMO~\cite{kong2023accumo} and Elf~\cite{zhang2021elf} use edge offloading to distribute tasks to remote devices. 
Despite their contributions, these solutions do not address fine-grained GPU kernel manipulation at runtime.
  
\noindent\textbf{GPU Kernel Scheduling and Manipulation. } 
{Several recent studies have explored fine-grained GPU kernel manipulation at runtime. 
Heimdall~\cite{yi2020heimdall} and dCUDA~\cite{ccuda} enable GPU time-sharing by partitioning applications into groups and executing them in distinct time slices, 
while cCUDA~\cite{ccuda} improves GPU concurrency through space sharing by splitting kernels into sub-kernels.  
PAAM~\cite{enright2024paam} and DART~\cite{xiang2019pipelined} prioritize kernel launches using CUDA streams based on predefined task criticality, 
with PAAM further optimizing both CPU/GPU prioritization for task chains and achieving state-of-the-art performance.   
However, none of these approaches considers priority, timing, and synchronization within a unified scheduling framework.
}

\noindent\textbf{Transparent GPU Management. }
Several recent efforts have focused on transparent GPU management.  
GPES~\cite{GPES} utilizes just-in-time binary code rewriting to insert SASS code into GPU kernels, enabling faster preemption.  
TGS~\cite{wu2023transparent} intercepts interactions between containers and GPUs to facilitate performance isolation and memory over-subscription.  
VCUDA~\cite{vcuda} employs API interception for GPU virtualization.   
GPUSync~\cite{gpusync} and BWLOCK++\cite{ali2017protecting} also adopt API interception, targeting deterministic execution. 
However, these methods focus primarily on kernel launch timing and are not designed for real-time autonomous systems.

\vspace{-6pt}
\section{ Conclusion }
 In this paper, we presented \systemname, an urgency-centric, transparent task scheduling system for closed-source, multi-task-chain applications with mixed real-time constraints and hybrid CPU-GPU usage. Our results show that \systemname outperforms the state-of-the-art in overall deadline miss rates under realistic autonomous driving scenarios. We believe \systemname will enable more responsive and adaptive autonomous driving systems in the future. Further, we believe \systemname's scheduling strategies could be extended to similar CPS applications with predictable workflows and dynamic data, offering insights to optimize mobile GPU tasks.

\begin{acks}
This work was supported by the National Natural Science Foundation of China (No. 62332016) and the Key Research Program of Frontier Sciences, CAS (No. ZDBS-LY-JSC001). 
Yanyong Zhang is the corresponding author.
\end{acks}

\bibliographystyle{ACM-Reference-Format}
\bibliography{arxiv}
\end{sloppypar}
\end{document}